\documentclass[10pt,twocolumn,twoside]{IEEEtran}

\usepackage[utf8]{inputenc} 
\usepackage[T1]{fontenc}    
\usepackage{hyperref}       
\usepackage{url}            
\usepackage{booktabs}       
\usepackage{amsfonts, amsmath}       
\usepackage{nicefrac}       
\usepackage{microtype}      

\usepackage{color,graphicx,amssymb,epsfig,mathrsfs,multirow,array,amsthm,cite}
\usepackage{wrapfig}
\usepackage{makecell}
\usepackage[caption=false,font=footnotesize]{subfig}
\usepackage[normalem]{ulem}
\usepackage{algorithm2e, setspace}

\usepackage{capt-of}

\newtheorem{theorem}{Theorem}[]
\everymath{\textstyle}
\let\displaystyle\textstyle
\usepackage[flushleft]{threeparttable}

\newcommand{\blue}[1]{{#1}} 
\newcommand{\squeezeup}{\vspace{-0.3cm}}
\newcommand{\MI}{\mathrm{MI}}
\newcommand{\I}{\mathrm{I}}

\IEEEoverridecommandlockouts

\begin{document}
\title{Mutual Information in Frequency and its Application to Measure Cross-Frequency Coupling in Epilepsy}
\author{Rakesh Malladi,~\IEEEmembership{Member,~IEEE,} Don H Johnson,~\IEEEmembership{Fellow, IEEE,}  Giridhar P Kalamangalam, \\ Nitin Tandon, and Behnaam Aazhang,~\IEEEmembership{Fellow,~IEEE} 
\thanks{This work is funded in part by grant 1406447 from National Science Foundation and Texas Instruments and was done at Rice University. A portion of this work was presented at Cosyne \cite{malladi2017a} and Asilomar \cite{malladi2017b}.}
\thanks{Rakesh Malladi is with LinkedIn Corporation, Sunnyvale, CA. Don H Johnson and Behnaam Aazhang are with the Department of Electrical and Computer Engineering, Rice University, Houston, TX. \blue{Giridhar P Kalamangalam is with Department of Neurology at University of Florida, Gainesville, FL.}  Nitin Tandon is with Department of Neurosurgery at University of Texas Health Center, Houston, TX. E-mail: malladirakesh@gmail.com, \{dhj, aaz\}@rice.edu, \blue{Giridhar.Kalamangalam@neurology.ufl.edu}, Nitin.Tandon@uth.tmc.edu.}
}

\maketitle

\begin{abstract}
We define a metric, mutual information in frequency (MI-in-frequency), to detect and quantify the statistical dependence between different frequency components in the data, referred to as cross-frequency coupling and apply it to electrophysiological recordings from the brain to infer cross-frequency coupling. The current metrics used to quantify the cross-frequency coupling in neuroscience cannot detect if two frequency components in non-Gaussian brain recordings are statistically independent or not. Our MI-in-frequency metric, based on Shannon's mutual information between the Cram\'{e}r's representation of stochastic processes, overcomes this shortcoming and can detect statistical dependence in frequency between non-Gaussian signals. We then describe two data-driven estimators of MI-in-frequency: one based on kernel density estimation and the other based on the nearest neighbor algorithm and validate their performance on simulated data. We then use MI-in-frequency to estimate mutual information between two data streams that are dependent across time, without making any parametric model assumptions. Finally, we use the MI-in-frequency metric to investigate the cross-frequency coupling in seizure onset zone from electrocorticographic recordings during seizures. The inferred cross-frequency coupling characteristics are essential to optimize the spatial and spectral parameters of electrical stimulation based treatments of epilepsy.
\end{abstract}

\begin{IEEEkeywords}
Mutual information in frequency; dependent data; Cram\'{e}r's spectral representation; cross-frequency coupling; epilepsy; seizure onset zone.
\end{IEEEkeywords}

\section{Introduction}
\blue{Epilepsy is a very common neurological disorder affecting nearly $1\%$ of the world's population. Epilepsy is characterized by repeated, unprovoked seizures. Nearly a third of all epilepsy patients have medically refractory epilepsy (medication is not effective in these patients). For these patients, surgical resection of the seizure onset zone (SOZ) (the regions of the brain responsible for generating and sustaining seizure activity \cite{luders2006}) or electrical stimulation are possible treatment options. However, the efficacy of these treatments is variable and almost always never results in a cure \cite{rosenow2001, bergey2015}. There is tremendous interest in leveraging the recent advances in electrical stimulation \cite{sunderam2010} and optogenetics \cite{krook2015} to develop spatiotemporally specific approaches to treat epilepsy. A crucial step in this endeavor is to develop an understanding of the coupling between neuronal oscillations in different frequency bands during seizures. This coupling or statistical dependence across frequency components between signals is referred to as cross-frequency coupling (CFC) \cite{canolty2006, canolty2010}. Our main objective is to learn the dynamics of cross-frequency coupling during seizures in epilepsy patients from the electrocorticographic (ECoG) data.}

\blue{Elaborating the characteristics of epileptic seizures using cross-frequency coupling between ECoG data has been the focus of many papers. CFC has been used to predict the onset of seizure in \cite{alvarado2014} and detect epileptic seizures in \cite{edakawa2016}. CFC has also been used to localize the area for surgical resection in epilepsy patients \cite{guirgis2015, weiss2015, liu2016}. Variations in CFC from preictal (before a seizure) to ictal (during a seizure) to postictal (after a seizures) in epilepsy patients have been analyzed in \cite{zhou2016, zhang2017}. In addition, the CFC in interictal stages is compared with that around seizures in \cite{amiri2016, cotic2016, edakawa2016, frauscher2017}. In this paper, we study CFC within and between various regions inside the seizure onset zone to determine the dominant frequencies involved in seizures and to learn the variations in coupling strength between various spatial regions inside SOZ. The results from this study are crucial to optimize the spectral and spatial parameters of next generation epilepsy treatments.}

\blue{Cross-frequency coupling or dependence across frequencies in the data could be in a single recording or between recordings, not necessarily at the same frequency. Coherence can identify if two frequency components are statistically independent or not and quantify the dependence for linear, Gaussian processes \cite{faes2011}. There is no such equivalent metric for non-Gaussian signals. Since the time-series data recorded from the brain are neither linearly related nor Gaussian, neuroscientists typically use heuristic metrics that cannot identify if two frequency components are statistically independent or not and can only capture second-order dependencies. Some of the popular heuristics estimate the phase-amplitude, amplitude-amplitude, phase-phase coupling between the low and high frequency components in the electrophysiological recordings from brain \cite{canolty2006, onslow2011, aru2015, pascual2016}.} In fact, a recent review article on CFC metrics suggests the use of cross-frequency `correlation' instead of `coupling' to describe these heuristic CFC metrics \cite{aru2015}. Furthermore, a list of confounds affecting the current CFC metrics is provided in \cite{aru2015}. A more comprehensive metric that detects statistical independence and thereby, capture both linear and nonlinear dependencies, would be invaluable in determining how neuronal oscillations at various frequencies are involved in the computation, communication, and learning in the brain. \blue{Here we propose a new methodology or metric to estimate the cross-frequency coupling (CFC) in neuroscience that overcomes the challenges of the existing approaches and as a proof-of-concept, we infer CFC characteristics of epileptic seizures using our metric.}

\blue{Mutual information in frequency (MI-in-frequency), defined for linear Gaussian processes using coherence in \cite{brillinger2002, salvador2008}, can indeed be further developed into a general technique to estimate CFC. Inspired by prior work \cite{brillinger2007}, we define MI-in-frequency between two frequencies in a signal (or two signals) as the Shannon's mutual information (MI) between the Cram\'{e}r's spectral representations \cite{cramer1967, larson1979} of the two signals at the corresponding frequencies. Cram\'{e}r's spectral representation transforms a time-domain stochastic process into a stochastic process in the frequency domain, the samples of which can be estimated at each frequency from the time-domain data samples \cite{brillinger2001}. MI-in-frequency metric is equivalent to coherence measures for linear, Gaussian signals and can be thought of as `coherence' for non-Gaussian signals.} The MI-in-frequency metric is one of the three mutual information based metrics used in \cite{brillinger2007} to analyze linear relationships between seismic data and \cite{brillinger2007} is not focussed on defining a single metric to capture the statistical dependence across frequency. We extend this approach to define a single metric, MI-in-frequency, to capture statistical dependencies across frequency for both linear and nonlinear data and use it measure CFC in the brain. We then describe two data-driven algorithms -- one based on kernel density estimation (KDMIF) and the other based on nearest neighbor estimation (NNMIF) -- to estimate MI-in-frequency without assuming any parametric model of the data. We considered these two approaches since they outperformed other approaches in estimating MI from i.i.d. data and there is no clear winner between them \cite{khan2007, schaffernicht2010}. We also demonstrate the superiority of MI-in-frequency over existing CFC metrics by comparing against modulation index \cite{canolty2006, onslow2011}, a commonly used CFC metric, on simulated data.

\blue{In addition to estimating CFC between ECoG data, we} use MI-in-frequency to develop a data-driven estimator for mutual information (MI). Note that MI estimation is a solved problem if the data samples are i.i.d. \cite{wang2009} or are sampled from linear, Gaussian processes \cite{pinsker1960, brillinger2002, salvador2008, cover2012}. As mentioned earlier, real-world data is neither independent across time nor Gaussian and the underlying model is often unknown. Our data-driven MI estimation algorithm applies to dependent data, without making any parametric model assumptions. The key idea is to make the problem computationally tractable by focussing only on those frequencies in the two data streams that are statistically dependent, which are identified by MI-in-frequency metric. Our MI estimator converges to the true value for Gaussian models and we validate its performance on nonlinear models.

Finally, we apply the MI-in-frequency estimators to infer the cross-frequency coupling in the seizure onset zone (SOZ), by analyzing electrocorticographic (ECoG) data from the SOZ of \blue{$9$ patients with medial temporal lobe epilepsy in whom a total of 25 seizures were recorded}. We investigate the dynamics of CFC in preictal, ictal and postictal periods within one SOZ electrode and between electrodes in different regions in the SOZ. We observe an increase in coupling in gamma and ripple high-frequency oscillations during seizures, with the largest increase within a SOZ electrode and a very small increase between electrodes in different regions inside SOZ. In addition, low-frequency coupling and linear interactions between SOZ electrodes also increase during the postictal state.

\section{Cram\'{e}r's Spectral Representation of Stochastic Processes} \label{sec:spec_rep}
Consider a stochastic processes $X\left(t\right), t \in \mathbb{R}$. Let $S_X\left(\nu\right)$ for $\nu \in \mathbb{R}$ be the spectral distribution function of $X$ and $s_X\left(\nu\right)$, its power spectral density, if it exists. Two basic spectral representations are associated with the stochastic process $X\left(t\right)$ -  power spectral distribution and Cram\'{e}r's representation \cite{cramer1967,larson1979}. The Cram\'{e}r's representation of $X\left(t\right)$ and its key properties are stated in the following theorem.
\begin{theorem} \label{theorem1}
\blue{(page 380 in \cite{larson1979})} Let $X\left(t\right)$ be a second order  stationary, mean-square continuous and zero mean stochastic process. Then there exists a complex-valued, finite-variance, orthogonal increment process $\widetilde{X}\left(\nu \right)$ in the frequency domain $\nu \in \mathbb{R}$, such that
\begin{align*} 
X\left(t\right) = & \int\limits_{-\infty}^{\infty} e^{j2\pi \nu t} d\widetilde{X}\left(\nu \right), \nonumber \\
 \text{with} \: \mathbb{E}\left[d\widetilde{X}\left(\nu \right) \right]  =  0, \: & \text{and} \: \mathbb{E}\left[|d\widetilde{X}\left(\nu \right)|^2\right]  =  dS_X\left(\nu\right).
\end{align*} 
\end{theorem} 
The process $\widetilde{X}\left(\nu\right) =  \widetilde{X}_R\left(\nu\right)+j\widetilde{X}_I\left( \nu\right)$ satisfying the above theorem is the spectral process or the Cram\'{e}r's representation of $X\left(t\right)$. $d\widetilde{X}\left(\nu\right)$ is the complex random variable representing the amplitude of oscillation in the interval from $\nu$ to $\nu + d\nu$ in $X\left(t\right)$. The integral in Theorem~\ref{theorem1} is a Fourier-Stieltjes integral. Intuitively, Theorem~\ref{theorem1}  decomposes $X\left(t\right)$ into an orthogonal increment complex process in the frequency domain. Furthermore, if the $X\left(t\right)$ is real-valued, then $\widetilde{X}\big(-\nu\big) = \widetilde{X}^{\star}\big(\nu\big), \mathbb{E}\big[d\widetilde{X}_R\big(\nu\big)d\widetilde{X}_I\big(\nu\big) \big] = 0$,  and
\begin{align} \label{eq2}
\mathbb{E}\big[\big(d\widetilde{X}_R\left(\nu\right) \big)^2 \big]=\mathbb{E}\big[\big(d\widetilde{X}_I\left(\nu\right) \big)^2 \big] = \frac{1}{2} dS_X \left( \nu \right).
\end{align} 
We have the following theorem for the special case of a real-valued Gaussian process $X\left(t\right)$.
\begin{theorem}\label{theorem2}
\blue{(page 385 in \cite{larson1979})} Let $X\left(t\right)$ be a real-valued stationary, mean-square continuous Gaussian process with zero mean and power spectral distribution function $S_X\left(\nu\right), \nu \in \mathbb{R}$. Then the real and imaginary parts of its spectral process $\widetilde{X}_R\left(\nu\right)$ and $\widetilde{X}_I\left(\nu\right)$ are zero mean, mutually independent, identically distributed Gaussian processes satisfying \eqref{eq2}.
\end{theorem}

\textit{Example:} Consider the zero mean stationary Gaussian process $X\left(t\right) = A\cos \left(2\pi \nu_0 t + \Theta\right)$, where $A$ is Rayleigh random variable with parameter $\sigma_A$ that is independent of $\Theta$, which is uniform in $\left[0,2\pi\right)$. The increments of  the spectral process of  $X\left(t\right)$ are all zero, except at $\nu=\pm\nu_0$, where the increment is $\frac{A}{2}\exp\left(\pm j\Theta\right)$ \cite{larson1979}. This implies that the sample path of the real part of spectral process $\widetilde{X}\left(\nu\right)$ has two jumps of same magnitude and direction at frequencies $\pm \nu_0$, while that of the imaginary part has two jumps of same magnitude, but opposite directions at $\pm \nu_0$. The magnitude of the jump at $\nu_0$ in the real and imaginary parts is $\frac{A}{2}\cos \Theta$ and $\frac{A}{2}\sin  \Theta$ respectively, both of which are Gaussian random variables with mean zero and variance $\frac{1}{2}\sigma_A^2$. This spectral process is intuitive because we know $X\left(t\right)$ has all its energy only at frequencies $\pm \nu_0$ and the variance of the increments of the spectral process $d\widetilde{X}\left(\nu\right)$ is equal to the differential power spectral distribution of $X\left(t\right)$ which is nonzero only at $\pm \nu_0$. We therefore expect all sample paths of the random process $\widetilde{X}\left(\nu\right)$ with non-zero probability to be constant, except for jumps at $\pm \nu_0$.

\blue{Note that if the process is wide sense-stationary and Gaussian, then power spectral distribution would have all the information about the process and its relationship with Cram\'{e}r's representation is given by Theorem~\ref{theorem2}. Otherwise, power spectral distribution only captures the second-order dependencies in the process. Since ECoG signals are not Gaussian, we use Cram\'{e}r's representation to transform a time-domain stochastic process into a stochastic process in the frequency domain.}

\section{Mutual Information in Frequency}
We first define MI between frequencies within a process and between two processes in continuous time. We then extend this definition to discrete-time stochastic processes. Consider $d\widetilde{X}\left(\nu_i\right)$ and $d\widetilde{Y}\left(\nu_j\right)$, the increments of spectral processes or the Cram\'{e}r's representation of $X(t)$ and $Y(t)$ at frequencies $\nu_i$ and $\nu_j$ respectively. Let the joint probability density of the four dimensional random vector of the real and imaginary parts of $d\widetilde{X}\left(\nu_i\right)$ and $d\widetilde{Y}\left(\nu_j\right)$ be denoted by $\mathrm{P}\big( d\widetilde{X}_R\left( \nu_i\right),d\widetilde{X}_I\left( \nu_i\right) ,d\widetilde{Y}_R\left( \nu_j\right),d\widetilde{Y}_I\left( \nu_j\right) \big)$. The corresponding two-dimensional marginal densities are $\mathrm{P}\big(d\widetilde{X}_R\left( \nu_i\right),d\widetilde{X}_I\left( \nu_i\right)\big)$, $\mathrm{P}\big(d\widetilde{Y}_R\left( \nu_j\right),d\widetilde{Y}_I\left( \nu_j\right)\big)$. The MI-in-frequency between $X\left(t\right)$ at $\nu_i$ and $Y\left(t\right)$ at $\nu_j$ is defined as
\begin{align} \label{MI_freq_def}
&\MI_{XY}\left(\nu_i,\nu_j\right) \nonumber \\
& = \I\big(\big\{d\widetilde{X}_R\left( \nu_i\right),d\widetilde{X}_I\left( \nu_i\right) \big\}; \big\{d\widetilde{Y}_R\left( \nu_j\right),d\widetilde{Y}_I\left( \nu_j\right) \big\} \big), \nonumber \\
& = \mathbb{E}\left\{ \log \frac{\mathrm{P}\big( d\widetilde{X}_R\left( \nu_i\right),d\widetilde{X}_I\left( \nu_i\right) ,d\widetilde{Y}_R\left( \nu_j\right),d\widetilde{Y}_I\left( \nu_j\right) \big)}{\mathrm{P}\big(d\widetilde{X}_R\left( \nu_i\right),d\widetilde{X}_I\left( \nu_i\right)\big) \mathrm{P}\big(d\widetilde{Y}_R\left( \nu_j\right),d\widetilde{Y}_I\left( \nu_j\right)\big)} \right\},
\end{align} 
where $\I\left(\left\{\cdot,\cdot\right\};\left\{\cdot,\cdot\right\} \right)$ is the standard mutual information between two pairs of two dimensional  real-valued random vectors \cite{cover2012}. The MI between two different frequencies $\nu_i$, $\nu_j$ in the same process $Y\left(t\right)$ is similarly defined as
\begin{align}\label{MI_freq_def_one_proc}
\!\!\MI_{YY}\!\big(\nu_i,\nu_j\big) \!\! = \!\! \I\big(\!\big\{\!d\widetilde{Y}_R\big( \nu_i\!\big),\!d\widetilde{Y}_I\big( \nu_i\!\big)\! \big\}\!;\! \big\{\!d\widetilde{Y}_R\big( \nu_j\!\big),\!d\widetilde{Y}_I\big( \nu_j\!\big)\! \big\}\! \big).
\end{align} 
The MI between the components of $Y$ at frequencies $\nu_i=\nu_j=\nu$, $\MI_{YY}\left(\nu,\nu\right)$, is $\infty$, a consequence of the fact that $\big[d\widetilde{Y}_R\left( \nu\right),d\widetilde{Y}_I\left( \nu\right) \big]$ is a continuous-valued random vector whose conditional differential entropy is not lower bounded. MI-in-frequency defined in \eqref{MI_freq_def}, \eqref{MI_freq_def_one_proc} is a non-negative number. \blue{If MI-in-frequency between two frequencies is zero, then they are independent and if not, MI-in-frequency is a measure of the statistical dependence between the two frequency components.} MI-in-frequency between two processes is not symmetric in general, i.e., $\MI_{XY}\left(\nu_i,\nu_j\right) \neq \MI_{XY}\left(\nu_j,\nu_i\right)$. However, it is symmetric within a process, i.e., $\MI_{YY}\left(\nu_i,\nu_j\right) = \MI_{YY}\left(\nu_j,\nu_i\right)$. 

\textit{Example:} Continuing with our example in section~\ref{sec:spec_rep}, let $X\left(t\right) = A\cos \left(2\pi \nu_0 t + \Theta\right)$ and $Y\left(t\right) = X\left(t\right)^2$. Then $d\widetilde{Y}\left(\nu\right)$ is zero except at $\nu=0$, where the spectral increment is $\frac{A^2}{2}$, and at $\nu=\pm2\nu_0$, where the increment is $\frac{A^2}{4}\exp\left(\pm j2\Theta\right)$. As a result, the frequency components at $\pm\nu_0$ in $X$ and at frequencies $\left\{0, \pm2\nu_0 \right\}$ in $Y$ are statistically dependent and hence the MI-in-frequency obtained from \eqref{MI_freq_def} at these frequency pairs will be positive. In addition, the frequency components in $Y$ at  $\nu \in \left\{0, \pm2\nu_0 \right\}$ are dependent and hence the MI-in-frequency within $Y$ at these frequencies will also be positive.

\subsection{Gaussian Inputs to LTI Filters} 
Let's now consider the special case where $X\left(t\right)$, a Gaussian process with power spectral density $s_X\left(\nu\right)$ serves as the input to a linear, time-invariant (LTI) filter \blue{with transfer function $H_1\left(\nu\right)$} and $Y\left(t\right)$ is output observed in additive colored noise \blue{(white noise $W\left(t\right)$ passed through a LTI filter with transfer function $H_2\left(\nu\right)$)}. The processes $X\left(t\right)$ and $Y\left(t\right)$ are related by 
\begin{align} \label{mvar_cont_eq}
y\left(t\right) = h_1\left(t\right) \ast x\left(t\right) + h_2\left(t\right) \ast w\left(t\right),
\end{align}
where $\ast$ denotes convolution operation, $x(t), y(t)$ and $w(t)$ are sample paths of $X\left(t\right)$, $Y\left(t\right)$ and $W\left(t\right)$ respectively. $W$ is a Gaussian process with power spectral density $s_W\left(\nu\right)$ and independent of $X$. $h_1(t)$ and $h_2(t)$ are continuous-time impulse responses of LTI filters, whose transfer functions are $H_1\left(\nu\right)$ and $H_2\left(\nu\right)$ respectively. Let $d\widetilde{X}\left(\nu\right)$, $d\widetilde{W}\left(\nu\right)$ and $d\widetilde{Y}\left(\nu\right)$ be the spectral process increments of the Gaussian processes $X$, $W$ and $Y$. We have from Theorem~\ref{theorem2},
\begin{align} \label{XW_spec_decomp}
\big[d\widetilde{X}_R\left(\nu\right),d\widetilde{X}_I\left(\nu\right) \big] \!\! & \sim \!\! \mathcal{N}\left(\mathbf{0},\frac{1}{2}s_X\left(\nu\right)\mathbf{I} \right), \nonumber \\
\big[d\widetilde{W}_R\left(\nu\right),d\widetilde{W}_I\left(\nu\right) \big] \!\! & \sim \!\! \mathcal{N}\left(\mathbf{0},\frac{1}{2}s_W\left(\nu\right)\mathbf{I} \right),
\end{align}
where $\mathcal{N}\left(\mathbf{\mu},\Sigma\right)$ represents Gaussian distribution with mean $\mathbf{\mu}$ and covariance $\Sigma$, $\mathbf{0}$ is the two element zero vector and $\mathbf{I}$ is the $2\times 2$ identity matrix. In addition, we can show for the model in \eqref{mvar_cont_eq} that 
\begin{align} \label{Y_spec_decomp}
d\widetilde{Y}\left(\nu\right) = H_1\left(\nu\right) d\widetilde{X}\left(\nu\right) + H_2\left(\nu\right) d\widetilde{W}\left(\nu\right).
\end{align}
\blue{The proof of \eqref{Y_spec_decomp} is in the appendix}. The MI-in-frequency defined in \eqref{MI_freq_def} is further simplified for the model in \eqref{mvar_cont_eq} using \eqref{XW_spec_decomp}, \eqref{Y_spec_decomp} and stated in the following theorem.
\begin{theorem} \label{theorem3}
For the model given in \eqref{mvar_cont_eq}, the MI between $X\left(t\right)$ at frequency $\nu_i$ and $Y\left(t\right)$ at frequency $\nu_j$ is zero, when $\nu_i \neq \nu_j$ and the MI between $X\left(t\right)$ and $Y\left(t\right)$ at frequency $\nu_i=\nu_j=\nu \neq 0$ is 
\begin{align} \label{theorem3eq}
\MI_{XY}\left(\nu,\nu\right) & = 2\times\I\big(\big\{d\widetilde{X}_R\left( \nu\right),d\widetilde{X}_I\left( \nu\right) \big\}; d\widetilde{Y}_R\left( \nu\right) \big) \nonumber \\
& =  \log \big(1 + \frac{|H_1\left(\nu\right)|^2s_X\left(\nu\right)}{|H_2\left(\nu\right)|^2s_W\left(\nu \right)} \big).
\end{align}
\end{theorem}
The proof of the above theorem is in the appendix. Note that at $\nu = 0$, the MI-in-frequency between $X$ and $Y$ is equal to $\I\big(\big\{d\widetilde{X}_R\left( \nu\right),d\widetilde{X}_I\left( \nu\right) \big\}; d\widetilde{Y}_R\left( \nu\right) \big)$, which is just half of the right hand side of \eqref{theorem3eq}. We intuitively expect different frequency components in the Gaussian input and its output from a linear system to be independent and Theorem~\ref{theorem3} confirms that the proposed definition of MI-in-frequency agrees with this intuition. In addition, the MI between $X$ and $Y$ is $\infty$ when $|H_2\left(\nu\right)| = 0$, since the components of $X$ and $Y$ at such $\nu$ are linearly related. The MI between two different frequencies in $Y\left(t\right)$, generated from \eqref{mvar_cont_eq}, is zero due to the linearity of the filters and Gaussian inputs. Furthermore, we can also show for the Gaussian processes $X$ and $Y$ related by \eqref{mvar_cont_eq} that MI-in-frequency is related to coherence $C_{XY}\left(\nu\right) \in \left[0,1 \right]$, by $\MI_{XY}\left(\nu,\nu\right)=-\log\left(1-C_{XY}\left(\nu\right)\right)$. The proof is in the appendix. This result implies MI-in-frequency between Gaussian processes related by \eqref{mvar_cont_eq} can be estimated with the  coherence. In addition, Theorem~\ref{theorem3} also shows that MI-in-frequency between Gaussian processes related by \eqref{mvar_cont_eq} can be estimated by estimating the mutual information between $\big[d\widetilde{X}_R\left( \nu_i\right),d\widetilde{X}_I\left( \nu_i\right) \big]$ and $d\widetilde{Y}_R\left( \nu_j\right)$, a three dimensional estimate as opposed to a four dimensional estimate in general.

\subsection{Discrete-time Stochastic Processes} 
We now extend the definition of MI-in-frequency between continuous-time stochastic processes in \eqref{MI_freq_def}, \eqref{MI_freq_def_one_proc} to discrete-time stochastic processes. In practice, we only have access to data samples from a real-valued, discrete-time stochastic process, sampled at a given Nyquist sampling frequency $F_s$. Sampled signals have periodic spectra, with a period equalling $F_s$. In addition, components in the process with frequencies in the range $\left[F_s/2,F_s\right]$ correspond to negative frequencies \cite{oppenheim1989}. Therefore, the actual frequency content in the signal is confined to $\left[0, F_s/2\right]$. We use normalized frequency $\lambda = \frac{\nu}{F_s} \in \left[0,0.5\right]$ to describe the frequency axis in case of discrete-time stochastic processes, instead of $\nu$ which was used for continuous-time stochastic processes. The MI-in-frequency between discrete-time processes is therefore obtained by replacing $\nu_i, \nu_j$ by the normalized frequencies $\lambda_1,\lambda_2 \in \left[0,0.5\right]$ in \eqref{MI_freq_def}, \eqref{MI_freq_def_one_proc}. Multivariate autoregressive models,  commonly used to model electro-physiological signals recorded from brain \cite{faes2011,pereda2005}, are a special case of the discrete-time equivalent of \eqref{mvar_cont_eq}. The analytic expression for MI at frequency $\lambda$ for such discrete-time Gaussian processes is therefore similarly obtained by replacing the frequencies $\nu$ by $\lambda$ in \eqref{theorem3eq}, which is also equal to $-\log\left(1-C_{XY}\left(\lambda\right)\right)$. This shows that for the special case of discrete-time Gaussian processes, MI-in-frequency metric is equivalent to coherence and the definitions in \cite{brillinger2002,salvador2008}.

\section{Data-Driven Estimation of MI-in-frequency} \label{sec:mif_est}
We describe two data-driven estimators--a kernel density based (KDMIF) and a nearest neighbor based (NNMIF) estimator to estimate MI-in-frequency, $\widehat{\MI}_{XY}\left(\lambda_i, \lambda_j \right)$, between $\lambda_i$ component of $X$ and $\lambda_j$ component of $Y$. The input to both these algorithms are the $N$ samples of $X$ and $Y$. The first step in both KDMIF and NNMIF estimators involves estimating the samples of spectral process increments $d\widetilde{X}\left(\lambda_i\right)$ and $d\widetilde{Y}\left(\lambda_j\right)$, of $X$ at $\lambda_i$ and of $Y$ at $\lambda_j$ respectively. In the second step, the KDMIF estimator uses the kernel density based MI estimator \cite{wang2009, scott2015}, whereas NNMIF estimator uses the k-nearest neighbor based MI estimator \cite{wang2009, kraskov2004} to estimate MI from the samples of spectral process increments, $d\widetilde{X}\left(\lambda_i\right)$ and $d\widetilde{Y}\left(\lambda_j\right)$.

\subsection{Kernel Density Based MI-in-frequency (KDMIF) Estimator} \label{sec:KDMIF}
\subsubsection{Estimation of Samples of Spectral Process Increments} \label{sec:kdmif_est_samples}
The first step of the algorithm is estimating the samples of spectral process increments of $X$ and $Y$ from $N$ dependent data samples. We assume there is a finite memory in both these processes and choose a value for a parameter $N_f$, which is much larger than the length of dependence or memory in the data and determines the frequency resolution of our MI-in-frequency estimates. We assume data in different windows are independent of each other. Ideally, consecutive windows should be separated to ensure no dependence across windows and avoid the dependence across the window boundaries, but our simulation results demonstrate that not separating the windows doesn't affect performance significantly. $N$ samples of $X$ are split into $N_s$ non-overlapping windows with $N_f = \frac{N}{N_s}$ data points in each window. Let us denote the samples in $l^{th}$ window of $X$ and $Y$ respectively by two $N_f$ element one-dimensional vectors, $\mathbf{x}^{l}$ and $\mathbf{y}^{l}$, for $l=1,2,\cdots,N_s$.

Let us now focus on estimating samples of the random variable $d\widetilde{X}\left(\lambda_i\right)$. Let $\mathcal{F}\left\{\mathbf{x}^l\right\}\left(\alpha \right)$ denote the discrete-time Fourier transform (DTFT) of $\mathbf{x}^{l}$ at normalized frequency $\alpha$. For $\lambda_i = \frac{i}{N_f} \in \left[0,1\right] \text{and} \: i \in \left[0, N_f-1 \right]$, let us define $d\widetilde{x}^{l}\left(\lambda_i\right)$ and integrated Fourier spectrum, $\widetilde{x}^{l}\left(\lambda_i\right)$, by
\begin{align} \label{spec_proc_inc_est}
\!\!d\widetilde{x}^{l}\left(\lambda_i\right) = \mathcal{F}\left\{\mathbf{x}^l \right\}\left(\lambda_i \right) \:  \text{and} \: \widetilde{x}^{l}\left(\lambda_i\right) = \sum\limits_{m=0}^{i} \mathcal{F}\left\{\mathbf{x}^l \right\}\left(\lambda_m \right).
\end{align}
It is stated in \cite{brillinger2001} that the random variable for which $\widetilde{x}^{l}\left(\lambda_i\right)$ is just one realization, tends to the spectral process of $X$ at $\lambda_i$ in mean of order $\gamma$, for any $\gamma > 0$, as the number of samples goes to infinity and assuming the underlying distribution is stationary and satisfies a mixing assumption. Also, $d\widetilde{x}^{l}\left(\lambda_i\right)$, which is the increment in $\widetilde{x}^l\left(\lambda_i\right)$ between $\lambda_i$ and $\lambda_i+d\lambda$, is just the DTFT of the samples in window $l$. Calculating the DTFT with the FFT for each of the $N_s$ windows separately yields an $N_f \times N_s$ matrix, whose $i^\mathrm{th}$ row, $\mathbf{d\widetilde{x}}\left(\lambda_i\right) = \left[d\widetilde{x}^{1}\left(\lambda_i\right),d\widetilde{x}^{2}\left(\lambda_i\right),\cdots,d\widetilde{x}^{N_s}\left(\lambda_i\right)\right]$ is the complex-valued vector containing $N_s$ samples of  $d\widetilde{X}\left(\lambda_i\right)$, the spectral process increments of $X$ at $\lambda_i=\frac{i}{N_f}$. The $l^{th}$ element of $\mathbf{d\widetilde{x}}\left(\lambda_i\right)$, $d\widetilde{x}^{l}\left(\lambda_i\right) = d\widetilde{x}_R^{l}\left(\lambda_i\right) + i d\widetilde{x}_I^{l}\left(\lambda_i\right)$,  is a particular realization of $d\widetilde{X}\left(\lambda_i\right)$. A similar procedure is used to obtain the $N_s$ samples of the spectral process increments of $Y$ at $\lambda_j= \frac{j}{N_f},  j \in \left[0, N_f-1 \right]$ and the resulting samples are denoted by $\mathbf{d\widetilde{y}}\left(\lambda_j\right) = \left[d\widetilde{y}^{1}\left(\lambda_j\right),d\widetilde{y}^{2}\left(\lambda_j\right),\cdots,d\widetilde{y}^{N_s}\left(\lambda_j\right)\right]$.

\subsubsection{Estimating MI-in-frequency} \label{sec:kdmif_est}
The MI-in-frequency estimate is now obtained from the $N_s$ samples, $\left(d\widetilde{x}_R^{l}\big(\lambda_i\big), d\widetilde{x}_I^{l}\big(\lambda_i\big) \right)$ and $\left(d\widetilde{y}_R^{l}\big(\lambda_j\big), d\widetilde{y}_I^{l}\big(\lambda_j\big) \right)$, for $l=1,2,\cdots,N_s$, using a kernel density based plug-in nonparametric estimator \cite{wang2009}. The $N_s$ data samples are split into $N_{tr}$ training and $N_{ts}$ test samples. The training data is used to estimate the four-dimensional joint probability density $\mathrm{P}\big(d\widetilde{X}_R\left(\lambda_i\right), d\widetilde{X}_I\left(\lambda_i\right),d\widetilde{Y}_R\left(\lambda_j\right), d\widetilde{Y}_I\left(\lambda_j\right)\big)$. The density is estimated using a kernel density estimator with Gaussian kernels, the optimal bandwidth matrix selected using smoothed cross-validation criterion \cite{scott2015} and implemented using `ks' package in R \cite{duong2007}. The joint density is marginalized to estimate the two-dimensional densities, $\mathrm{P}\big(d\widetilde{X}_R\left(\lambda_i\right), d\widetilde{X}_I\left(\lambda_i\right)\big)$ and $\mathrm{P}\big(d\widetilde{Y}_R\left(\lambda_j\right), d\widetilde{Y}_I\left(\lambda_j\right)\big)$, by recognizing that the bandwidth matrix for the two-dimensional marginal is the appropriate $2\times 2$ sub-matrix from the $4\times4$ bandwidth matrix of the joint density. The estimates of the joint and the marginal densities at the $N_{ts}$ test samples are plugged into the following equation \eqref{MI_plugin} to estimate MI-in-frequency.
\begin{align} \label{MI_plugin}
&\widehat{\MI}_{XY}\left(\lambda_i,\lambda_j\right) \nonumber \\
& = \frac{1}{N_{ts}} \sum\limits_{l} \log \frac{\mathrm{\widehat{P}}\big(d\widetilde{x}^l_R\left(\lambda_i\right), d\widetilde{x}^l_I\left(\lambda_i\right),d\widetilde{y}^l_R\left(\lambda_j\right), d\widetilde{y}^l_I\left(\lambda_j\right) \big)}{\mathrm{\widehat{P}}\big(d\widetilde{x}^l_R\left(\lambda_i\right), d\widetilde{x}^l_I\left(\lambda_i\right)\big)\mathrm{\widehat{P}}\big(d\widetilde{y}^l_R\left(\lambda_j\right), d\widetilde{y}^l_I\left(\lambda_j\right) \big)}.
\end{align}

\subsection{Nearest Neighbor Based MI-in-frequency (NNMIF) Estimator} \label{sec:NNMIF}
\subsubsection{Estimation of Samples of Spectral Process Increments} \label{sec:nnmif_est_samples}
The first step in the nearest neighbor based MI-in-frequency estimator is exactly same as that of KDMIF estimator. Following the steps described in section~\ref{sec:kdmif_est_samples}, we estimate $d\widetilde{x}^l\left(\lambda_i\right)$ and $d\widetilde{y}^l\left(\lambda_j\right)$, for $l= 1, 2, \cdots, N_s$,  the $N_s$ samples of the spectral process increments of $X$ at $\lambda_i$ and $Y$ at $\lambda_j$ respectively.

\subsubsection{Estimating MI-in-frequency}\label{sec:nnmif_est}
${\MI}_{XY}\left(\lambda_i, \lambda_j \right)$ is now estimated from $d\widetilde{x}^l\left(\lambda_i \right)\in \mathbb{R}^2$ and $d\widetilde{y}^l\left(\lambda_j \right)\in \mathbb{R}^2$, for $l=1, 2, \cdots, N_s$ using nearest neighbor based MI estimator \cite{kraskov2004}. We apply the first version of the algorithm in \cite{kraskov2004} to two-dimensional random variables $d\widetilde{X}\left(\lambda_i\right)$ and $d\widetilde{Y}\left(\lambda_j\right)$ to compute $\widehat{\MI}_{XY}\left(\lambda_i, \lambda_j \right)$. Consider the joint four dimensional space $\big(d\widetilde{X}\left(\lambda_i\right), d\widetilde{Y}\left(\lambda_j\right)\big) \in \mathbb{R}^4$. The distance between two data points with indices $l_1, l_2 \in \left[1, N_s \right]$ is calculated using the infinity norm, according to $\max \left\{ \|d\widetilde{x}^{l_1}\left(\lambda_i\right) - d\widetilde{x}^{l_2}\left(\lambda_i\right) \|, \|d\widetilde{y}^{l_1}\left(\lambda_j\right) - d\widetilde{y}^{l_2}\left(\lambda_j\right) \| \right\}$. Let $\epsilon_l$ denote the distance between the data sample $\big(d\widetilde{x}^{l}\left(\lambda_i\right), d\widetilde{y}^{i}\left(\lambda_j\right)\big)$ and its $K^{th}$ nearest neighbor, for $l =1, 2, \cdots, N_s$. We used $K=3$ in this paper \cite{khan2007}. Let $n_x^l$ and $n_y^l$ denote the number of samples of $d\widetilde{X}\left(\lambda_i\right)$ and $d\widetilde{Y}\left(\lambda_j\right)$ within an infinity norm ball of radius less than $\epsilon_l$ centered at $d\widetilde{x}^{l}\left(\lambda_i\right)$ and $d\widetilde{y}^{i}\left(\lambda_j\right)$ respectively. From \cite{kraskov2004}, the MI-in-frequency between $X$ and $Y$ at normalized frequencies $\lambda_i$ and $\lambda_j$ is given by
\begin{align} \label{eq_knn_est}
\widehat{\MI}_{XY}\left(\lambda_i,\lambda_j\right) & = \psi\left(K\right) + \psi\left(N_s\right) \nonumber \\
& - \frac{1}{N_s}\sum\limits_{l=1}^{N_s} \left(\psi\left(n_x^l + 1\right) + \psi\left(n_y^l + 1 \right) \right),
\end{align}
where $\psi\left(\cdot\right)$ is the Digamma function. 

\subsection{Significance Testing} \label{sec:stat_sig}
The statistical significance of the MI-in-frequency estimates obtained from both KDMIF and NNMIF estimators is now tested using the following procedure. We permute the samples in the vector $\mathbf{d\widetilde{x}}\left(\lambda_i\right)$ randomly and estimate the MI-in-frequency between the permuted vector and the $N_s$ samples of $d\widetilde{Y}\left(\lambda_j\right)$. Instead of adding random phase or permuting the phase time series, which are typically used to test the statistical significance of phase-amplitude coupling metrics \cite{dvorak2014}, we permute the samples of spectral process increments since our metric can detect coupling across phase and amplitude jointly. This process is repeated $N_p$ times to obtain $N_p$ permuted MI-in-frequency estimates, under the null hypothesis of independence. The permuted MI estimates will be almost zero since the permutations make the spectral processes almost independent. If the actual MI estimate, $\widehat{\MI}_{XY}\left(\lambda_i,\lambda_j\right)$, is judged larger than all the permuted $N_p$ estimates, then there is a statistically significant dependence between the processes at these frequencies.

\section{MI between Data with Temporal Dependencies} \label{sec:mi_est}
We now use MI-in-frequency to estimate mutual information between dependent data. The data-driven MI estimator, summarized in Algorithm~\ref{mi_est_algo}, takes in $N$ samples of $X$ and $Y$ as input and outputs the mutual information between $X$ and $Y$, $\hat{\I} \left(X; Y \right)$, by estimating $\widehat{\MI}_{XY}\left(\lambda_i, \lambda_j \right)$, where $\lambda_i = \frac{i}{N_f}, \lambda_j=\frac{j}{N_f}$, $\forall \left(i, j\right) \: \text{such that} \: i, j \in \left[0, N_f-1 \right]$.

\RestyleAlgo{ruled}
\begin{algorithm}[!htbp]
\setstretch{1.15}
\SetAlgoLined
\KwData{$\left(x\left[n\right], y\left[n\right] \right)$, for $x\left[n\right], y\left[n\right] \in \mathbb{R}, n\in[0,N-1].$}
\KwResult{$\hat{\I} \left(X; Y \right)$} 
\SetKw{Kw}{Algorithm:}
\Kw
\begin{enumerate}
\item[\textit{A)}] Estimate $\widehat{\MI}_{XY}\left(\lambda_i, \lambda_j \right)$ at all possible pairs $\left(\lambda_i,\! \lambda_j \right)$, using either the KDMIF or the NNMIF estimator. Identify the sets $\Lambda_x, \Lambda_y$, such that for each $\lambda_{i_p} \in \Lambda_x$ there exists a $\lambda_{j_q} \in \Lambda_y$ such that $\widehat{\MI}_{XY}\left(\lambda_{i_p}, \lambda_{j_q} \right)$ is statistically significant and vice-versa. Let $P, Q$ respectively denote the cardinality of $\Lambda_x, \Lambda_y$.
\item[\textit{B)}] Let $d\widetilde{X}\left(\Lambda_x \right) = \big[d\widetilde{X}\left(\lambda_{i_1} \right), \cdots, d\widetilde{X}\left(\lambda_{i_P} \right) \big] \in \mathbb{R}^{2P}$, $d\widetilde{Y}\left(\Lambda_y \right) = \big[d\widetilde{Y}\left(\lambda_{j_1} \right), \cdots, d\widetilde{Y}\left(\lambda_{j_Q} \right) \big] \in \mathbb{R}^{2Q}$. The mutual information between $X$ and $Y$ is given by
\begin{align*}
\hat{\I} \left(X; Y \right) = \frac{1}{\max(P,Q)} \hat{\I} \left(d\widetilde{X}\left(\Lambda_x \right);  d\widetilde{Y}\left(\Lambda_y \right)\right),
\end{align*}
where the right hand side is estimated from $N_s$ i.i.d. samples using any nonparametric MI estimator \cite{wang2009}.
\end{enumerate}
\caption{Mutual Information Estimator\vspace{.05cm}}\label{mi_est_algo}
\end{algorithm}

\subsection{Identifying Coupled Frequencies} \label{subsec:coupled_freq}
The first step in our MI estimator involves estimating the MI-in-frequency, $\widehat{\MI}_{XY}\left(\lambda_i, \lambda_j \right)$, between $\lambda_i= \frac{i}{N_f}$ frequency component in $X$ and $\lambda_j=\frac{j}{N_f}$ component in $Y$, for all $ \left(i, j\right) \: \text{such that} \: i, j \in \left[0, N_f-1 \right]$ using either the KDMIF (section~\ref{sec:KDMIF}) or the NNMIF (section~\ref{sec:NNMIF}) algorithms. Statistical significance of the resulting estimates is assessed using the procedure described in section~\ref{sec:stat_sig}. The resultant MI-in-frequency estimates across all frequency pairs can be graphically visualized by plotting the statistically significant MI-in-frequency estimates on a two-dimensional image grid, whose rows and columns correspond to frequencies of $X$ and $Y$ respectively. Let $\Lambda_x$ and $\Lambda_y$ respectively denote the set of frequency components of $X$ and $Y$, such that for each $ \lambda_{i_p} \in \Lambda_x$, there exists at least one $ \lambda_{j_q} \in \Lambda_y$ for which $ \widehat{\MI}_{XY}\left( \lambda_{i_p}, \lambda_{j_q}\right)$ is statistically significant and vice-versa. 

\subsection{Estimating Mutual Information}
The final step in our algorithm estimates MI between the spectral process increments of $X$ and $Y$ at frequencies in $\Lambda_x$ and $\Lambda_y$ respectively. With $P, Q$ denoting the cardinality of $\Lambda_x, \Lambda_y$ respectively, let $d\widetilde{X}\left(\Lambda_x \right)$ and $d\widetilde{Y}\left(\Lambda_y \right)$ denote the $2P$ and $2Q$-dimensional random vector comprising the spectral process increments of $X$, $Y$ at all frequencies in $\Lambda_x$ and $\Lambda_y$ respectively. We already computed $N_s$ i.i.d. samples of these two random vectors to estimate MI-in-frequency estimates in the previous step of this algorithm. The desired MI estimate is computed from the mutual information between $d\widetilde{X}\left(\Lambda_x \right)$ and $d\widetilde{Y}\left(\Lambda_y \right)$, which is estimated using the k-nearest neighbor based estimator developed in \cite{kraskov2004}, according to
\begin{equation}\label{eq:mi_est}
\hat{\I} \left(X; Y \right) = \frac{1}{\max\left(P,Q\right)} \hat{\I} \left(d\widetilde{X}\left(\Lambda_x \right);  d\widetilde{Y}\left(\Lambda_y \right)\right).
\end{equation}

The MI estimator in \eqref{eq:mi_est} can be further simplified for discrete-time Gaussian processes. Without loss of generality, consider two Gaussian processes $X$ and $Y$, related by 
\begin{align} \label{eq:linear_model}
y[n] = h_1[n]*x[n] + h_2[n]*w[n],
\end{align} 
where $h_1[n], h_2[n]$ are linear time-invariant (LTI) filters and $W$ is white Gaussian noise independent of $X$. For the model in \eqref{eq:linear_model}, which is the discrete-time equivalent of \eqref{mvar_cont_eq}, the data-driven estimation in \eqref{eq:mi_est} can be further simplified to
\begin{align}\label{eq:linear_model_mi_est}
\hat{\I}\left(X;Y\right) =  \frac{1}{N_f}\sum\limits_{i=0}^{N_f/2} \widehat{\MI}_{XY}\left(\lambda_i ; \lambda_i\right), \: \text{where} \: \lambda_i = \frac{i}{N_f}.
\end{align}
This result is obtained because linear models do not introduce cross-frequency dependencies and because negative frequencies do not carry any extra information.  Furthermore, the relationship between the MI and the MI-in-frequency for two processes related by \eqref{eq:linear_model} is stated in the following theorem.
\begin{theorem} \label{theorem4}
Consider two discrete-time Gaussian stochastic processes $X$ and $Y$ related by \eqref{eq:linear_model}. The mutual information between these processes, a scalar, is given by 
\begin{align} \label{eq:linear_model_mi}
\I \left(X;Y\right) = \int\limits_{0}^{0.5} \MI_{XY}\left(\lambda,\lambda\right) d\lambda.
\end{align}
\end{theorem}
The proof of the above theorem is in the appendix. This theorem means that MI between two Gaussian processes over the entire time can be obtained by integrating the contribution from each frequency component. It is easy to see that the right hand side of \eqref{eq:linear_model_mi_est} is just the Riemann sum of the integral on the right hand side of \eqref{eq:linear_model_mi}, which converges to the true value as $N_f$ tends to infinity. This implies our MI estimator converges to the true value for discrete-time Gaussian processes.

Note that the MI estimation algorithm does not make any parametric assumptions on the underlying model between $X$ and $Y$. The computation of MI via \eqref{eq:mi_est} can be greatly simplified by clustering the frequencies in $\Lambda_x$ and $\Lambda_y$ into groups such that there are no significant dependencies across groups and using the chain rule of mutual information. In addition, if we observe after the first step that significant MI-in-frequency estimates occur only at $\left(\lambda_i, \lambda_i \right),\! \forall i \!\in\! \big[0, N_f-1\big]$, then the MI can be estimated using \eqref{eq:linear_model_mi_est}.

\blue{Finally, as we mentioned earlier, MI estimation between Gaussian processes is a solved problem in the sense that we can analytically compute it if the covariance of the Gaussian processes is known [3] and there are several estimators whose performance is thoroughly analyzed [4]. MI in frequency for Gaussian processes is analyzed by Brillinger \cite{brillinger2002}. In this paper, we extended Brillinger's work to define MI-in-frequency for any process. In the following section, we use simulated data to validate that the extensions we proposed to any process in this paper are still in agreement with the prior work on Gaussian processes and also work for non-Gaussian processes.}

\begin{figure*}[!t]
\centering
\subfloat[MI-in-frequency]{
\includegraphics[width=0.31\textwidth]{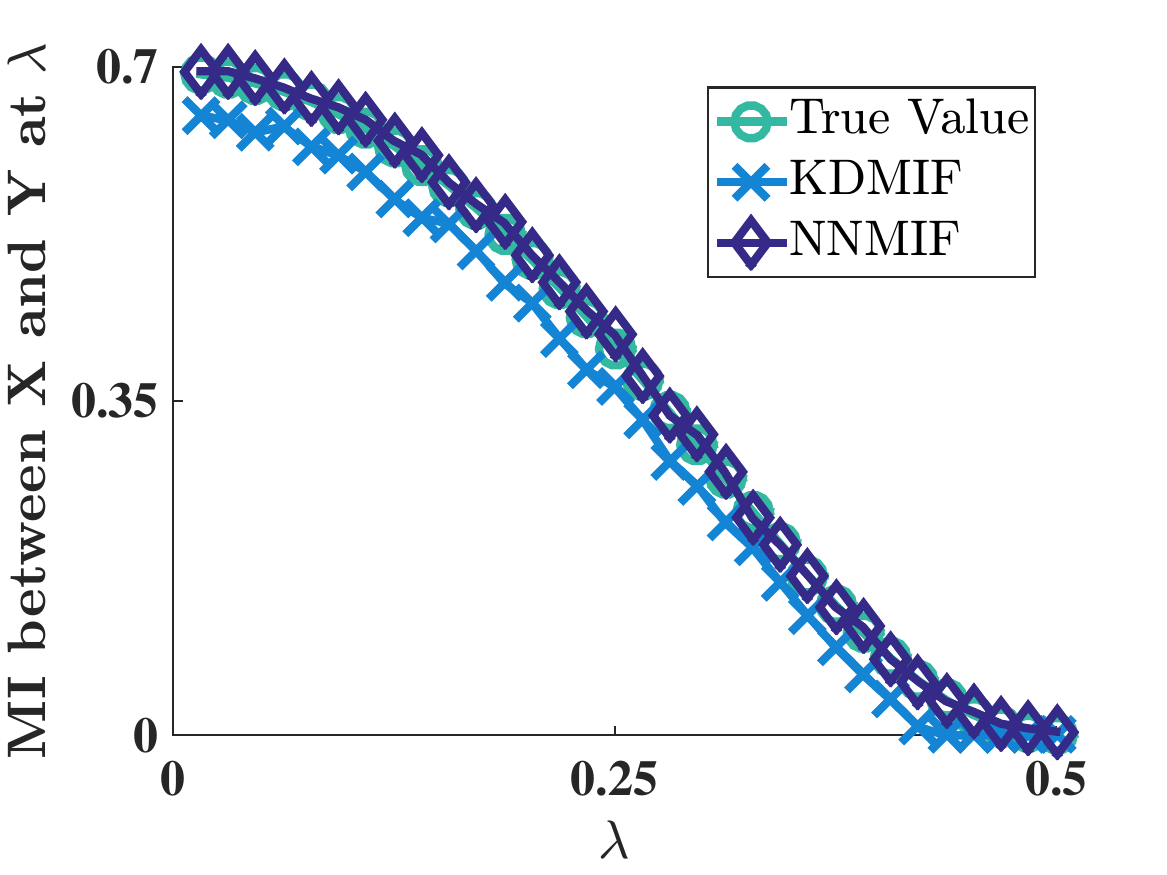} }
\hfill
\subfloat[Bias of the Estimators]{
\includegraphics[width=0.31\textwidth]{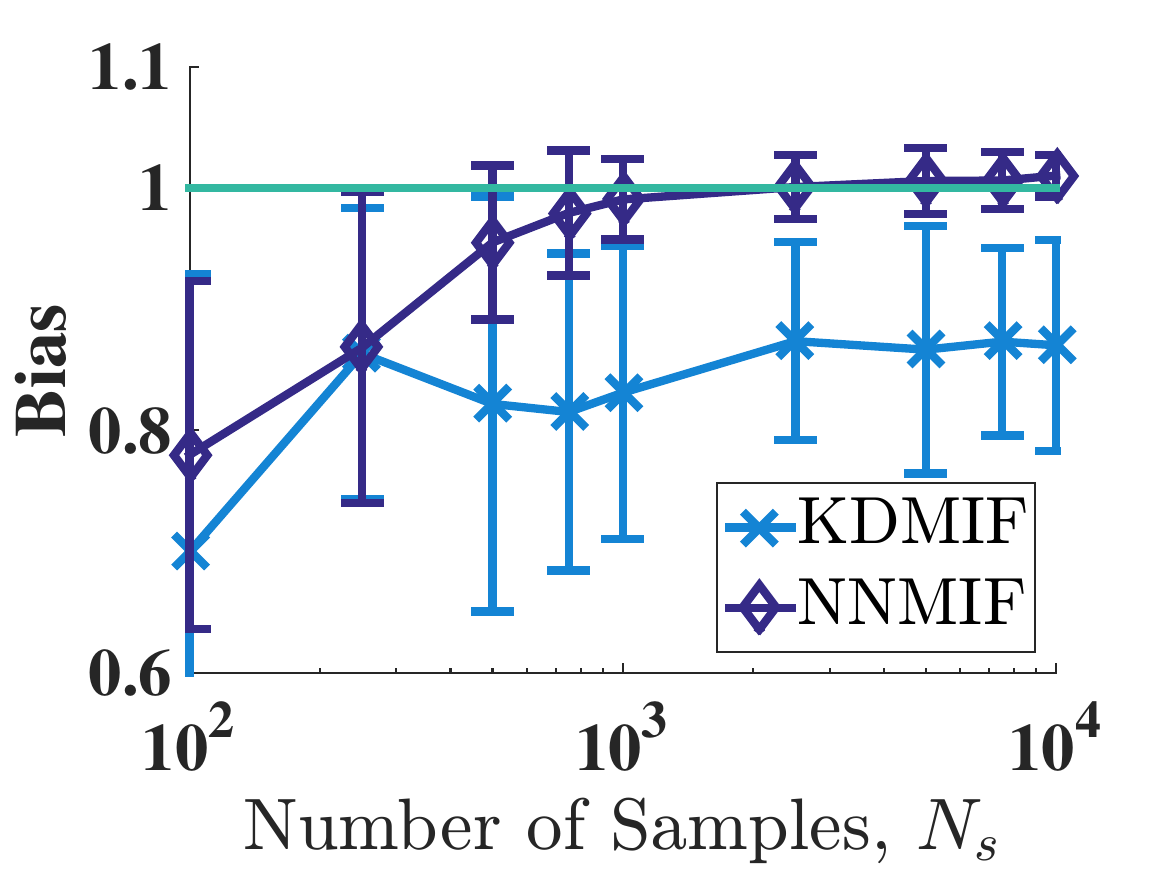} }
\hfill
\subfloat[Mutual Information]{
\includegraphics[width=0.31\textwidth]{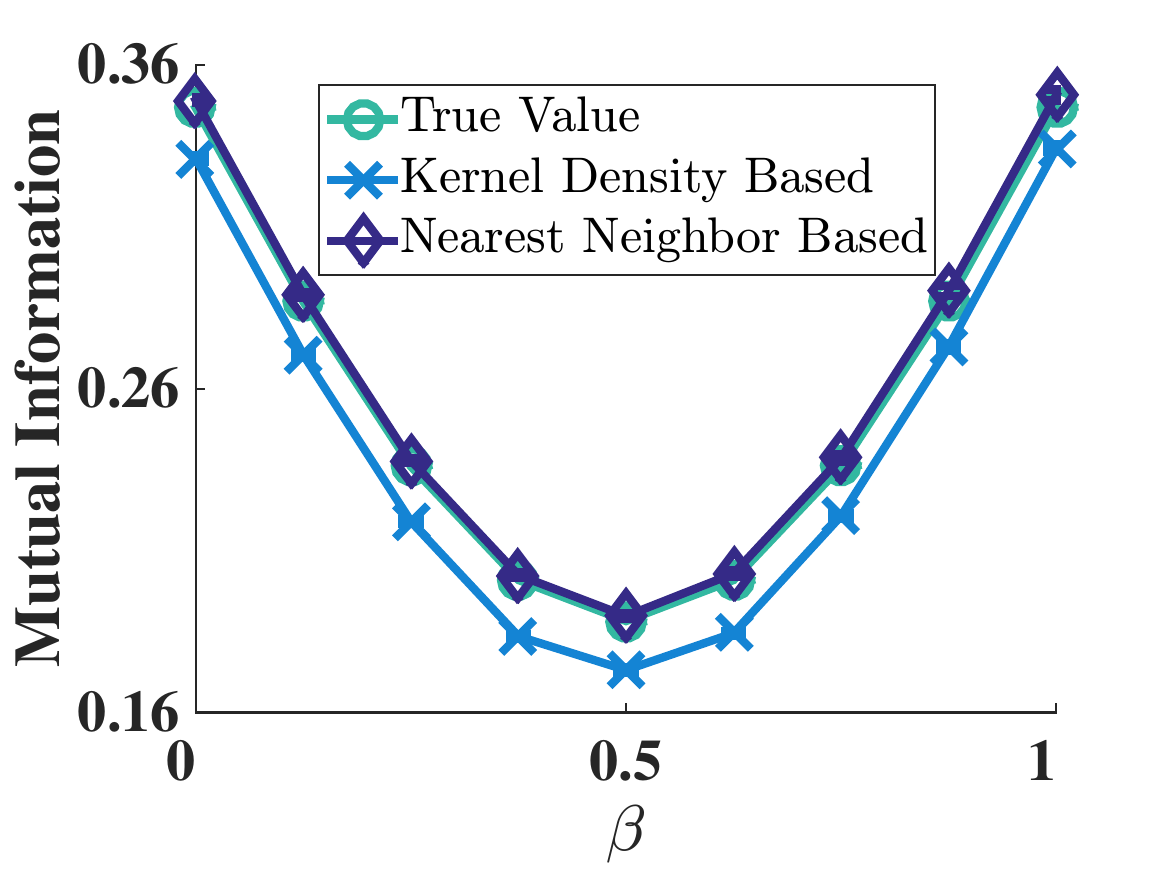} }
\caption{Comparing the performance of the kernel density based and nearest neighbor based estimators, KDMIF and NNMIF respectively, on simulated generated from \eqref{discrete_linear_eq} using a two-tap lowpass filter. In Fig.~\ref{Fig:lowpass}a, the MI-in-frequency estimates obtained from KDMIF and NNMIF estimators along with the true value of MI-in-frequency are plotted against the normalized  frequency $\lambda$ for $\beta=0.5$. Fig.~\ref{Fig:lowpass}b plots the bias (mean of the ratio of the estimate and the true value in the filter passband) against the number of data samples used for estimation for $\beta=0.5$. Fig.~\ref{Fig:lowpass}c plots the MI estimate between $X$ and $Y$ obtained from kernel density and nearest neighbor algorithms along with the true value of MI for $\beta \in \left[0, 1\right]$.}\label{Fig:lowpass}
\squeezeup
\end{figure*} 

\section{Performance Evaluation on Simulated Data}
The performance of the data-driven MI-in-frequency and mutual information estimators described in section~\ref{sec:mif_est} and section~\ref{sec:mi_est} respectively is validated on simulated data. The statistical significance of the estimates was assessed using the procedure described in section~\ref{sec:stat_sig}. In addition, we compare the performance of the MI-in-frequency estimators against modulation index \cite{canolty2006, canolty2010, aru2015}, a commonly used phase-amplitude coupling metric in neuroscience.

\subsection{Comparing the KDMIF and NNMIF Estimators}
Consider two stochastic processes $X$ and $Y$, where $X$ is a white Gaussian process with standard deviation $\sigma_x$ and $Y$ is obtained by
\begin{align}\label{discrete_linear_eq}
y[n] = h[n] \ast x[n] + w[n],
\end{align}
where $W$ is a white Gaussian process with standard deviation $\sigma_w$ that is independent of $X$ and $h[n]$ is a linear time-invariant filter. We compared the performance of the kernel density based and nearest neighbor based estimators by benchmarking the estimates against the true value of MI-in-frequency and the mutual information between $X$ and $Y$ for the model in \eqref{discrete_linear_eq}. We used two different filers: a two-tap low pass filter, $h[n] = \left[\beta, 1-\beta \right],$ for $\beta \in \left[0, 1\right]$ and a $33$-tap bandpass filter with passband in $\left[0.15, 0.35\right]$ normalized frequency range. We observed that modulation index, a popular CFC metric, was unable to correctly detect and quantify the strength of cross-frequency coupling for both these models. 
\begin{figure*}[!t]
\centering
\subfloat[MI-in-frequency]{
\includegraphics[width=0.31\textwidth]{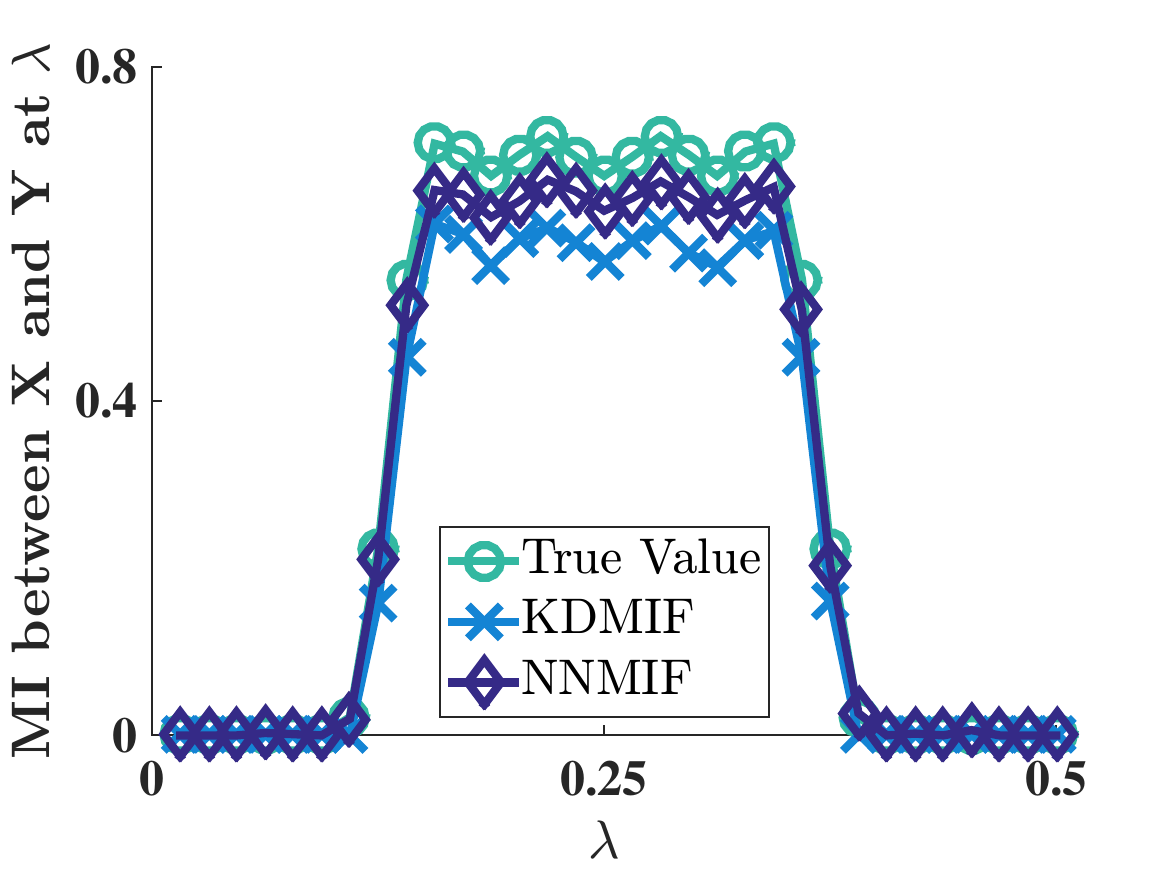} }
\hfill
\subfloat[Bias of the Estimators]{
\includegraphics[width=0.31\textwidth]{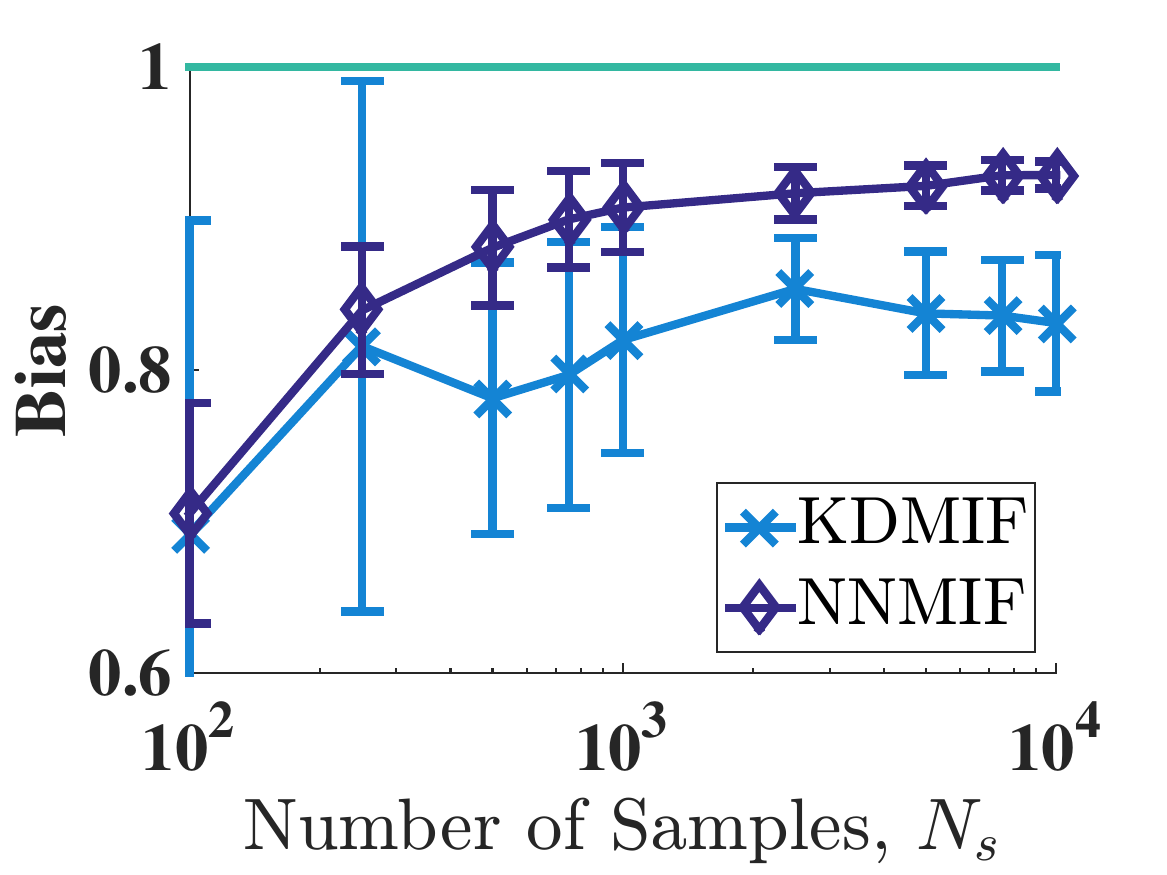} }
\hfill
\subfloat[Mutual Information]{
\includegraphics[width=0.31\textwidth]{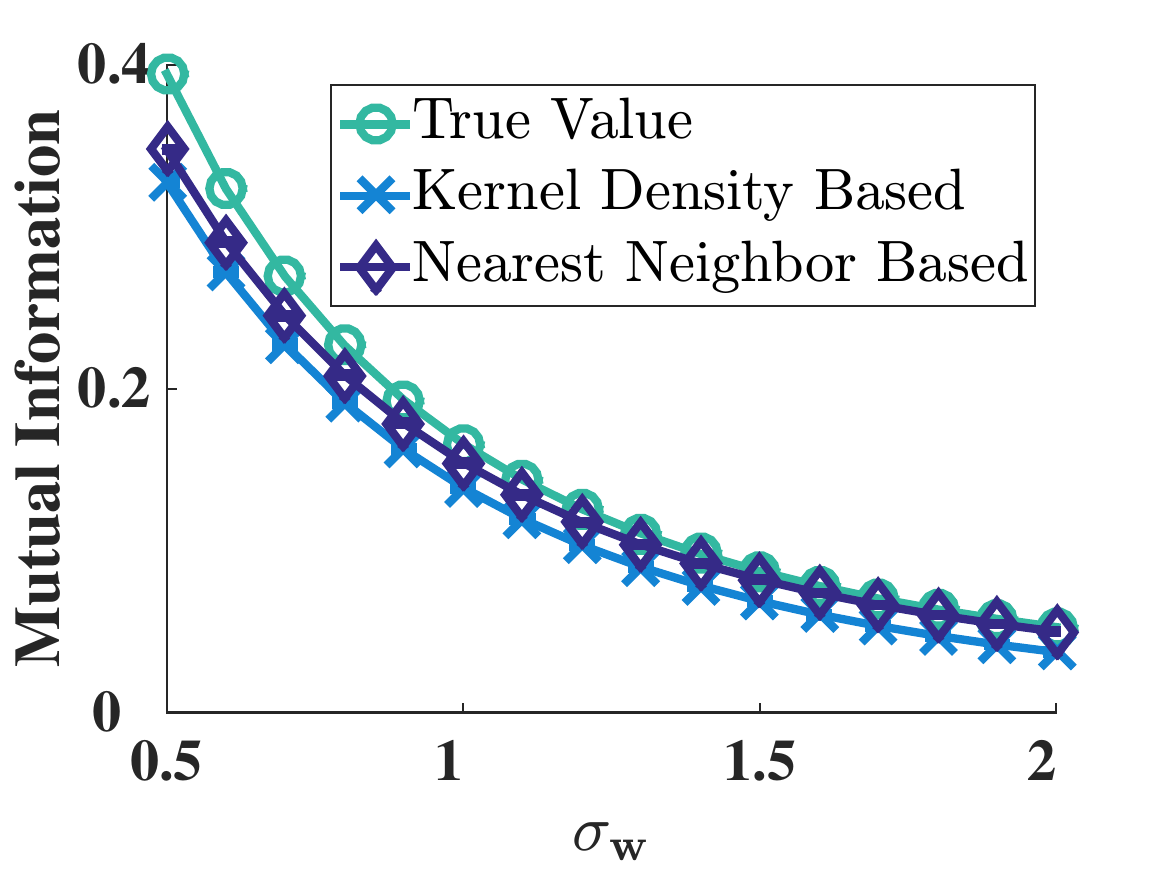} }
\caption{Comparing the performance of the kernel density based and nearest neighbor based estimators, KDMIF and NNMIF respectively, on simulated generated from \eqref{discrete_linear_eq} using a $33$-tap bandpass filter with passband in $\left[0.15, 0.35\right]$ normalized frequency. In Fig.~\ref{Fig:bandpass}a, the MI-in-frequency estimates obtained from KDMIF and NNMIF estimators along with the true value of MI-in-frequency are plotted against the normalized  frequency $\lambda$ for $\sigma_w = 1$. Fig.~\ref{Fig:bandpass}b plots the bias (mean of the ratio of the estimate and the true value in the filter passband) against the number of data samples used for estimation for $\sigma_w = 1$. Fig.~\ref{Fig:bandpass}c plots the plots the MI estimate between $X$ and $Y$ from kernel density and nearest neighbor algorithms along with the true value of MI for different values of $\sigma_w \in \left[0.5, 2\right]$.}\label{Fig:bandpass}
\squeezeup
\end{figure*}

\subsubsection{Lowpass Filter}
The samples of $X$ and $Y$ are generated from \eqref{discrete_linear_eq} with $\sigma_x = \sigma_w = 1$ and a lowpass filter with unit-impulse response $ \left[\beta, 1-\beta \right]$, for various values of $\beta \in \left[0, 1\right]$. The true value of MI-in-frequency at normalized frequency $\lambda \in \left[0, 0.5\right]$ is obtained substituting the parameters of this model in \eqref{theorem3eq} and is plotted in Fig.~\ref{Fig:lowpass}a for $\beta=0.5$. In addition, the MI-in-frequency estimated by the KDMIF and NNMIF algorithms from $N = 64\times 10^4$ data samples, with $N_f = 64, N_s = 10^4$ is also plotted in Fig.~\ref{Fig:lowpass}a. It is seen that the estimates from both algorithms follow the true value closely, without the knowledge of the underlying model. In addition, we evaluate the bias and the rate of convergence of both these algorithms as a function of $N_s$, with $N_f = 64$ in Fig.~\ref{Fig:lowpass}b. The bias is defined as the average value of the ratio of MI-in-frequency estimate and its true value in the passband of the lowpass filter. We observe that the NNMIF algorithm converges faster and has lower bias than the KDMIF algorithm. We now use both these algorithms to estimate the mutual information between $X$ and $Y$ for $\beta \in \left[0, 1\right]$. The analytical expression for the true value of MI\footnote{Note that for this particular model, mutual information is equal to the directed information from X to Y and the analytical expression is given in equation (18) in \cite{malladi2016}.} for this model is derived in \cite{malladi2016}. It is evident from Fig.~\ref{Fig:lowpass}c that the MI estimates obtained from the nearest neighbor based estimator is closer to the true value than those from the kernel density based estimator.

\subsubsection{Bandpass Filter}
The samples of $X$ are generated from a standard white Gaussian random process with $\sigma_x = 1$ and those of $Y$ are generated from \eqref{discrete_linear_eq} using a 33-tap finite-impulse-response bandpass filter with passband in $\left[0.15, 0.35 \right]$ normalized frequency range  for different values of noise standard deviation, $\sigma_w \in \left[0.5, 2\right]$. We used the kernel density and the nearest neighbor based algorithms to estimate the MI-in-frequency and the mutual information between $X$ and $Y$. The true value of MI-in-frequency is obtained from \eqref{theorem3eq} and of mutual information is numerically calculated using power spectral density (chapter 10 in \cite{cover2012}). It is clear from Fig.~\ref{Fig:bandpass}b that the nearest neighbor based algorithm converges to the true value faster than the kernel density based algorithm. The nearest neighbor based algorithm also provides more accurate estimates of both MI-in-frequency and mutual information between $X$ and $Y$, as evident from Fig.~\ref{Fig:bandpass}a, Fig~\ref{Fig:bandpass}c respectively. In addition, nearest neighbor based MI-in-frequency algorithm runs faster than kernel density based algorithm. We, therefore, conclude that the nearest neighbor based MI-in-frequency algorithm outperforms kernel density based algorithms and only depict the results obtained from nearest neighbor based algorithm in the remainder of the paper.

\begin{figure*}[!t]
\centering
\subfloat[MI-in-frequency]{
\includegraphics[width=.23\textwidth]{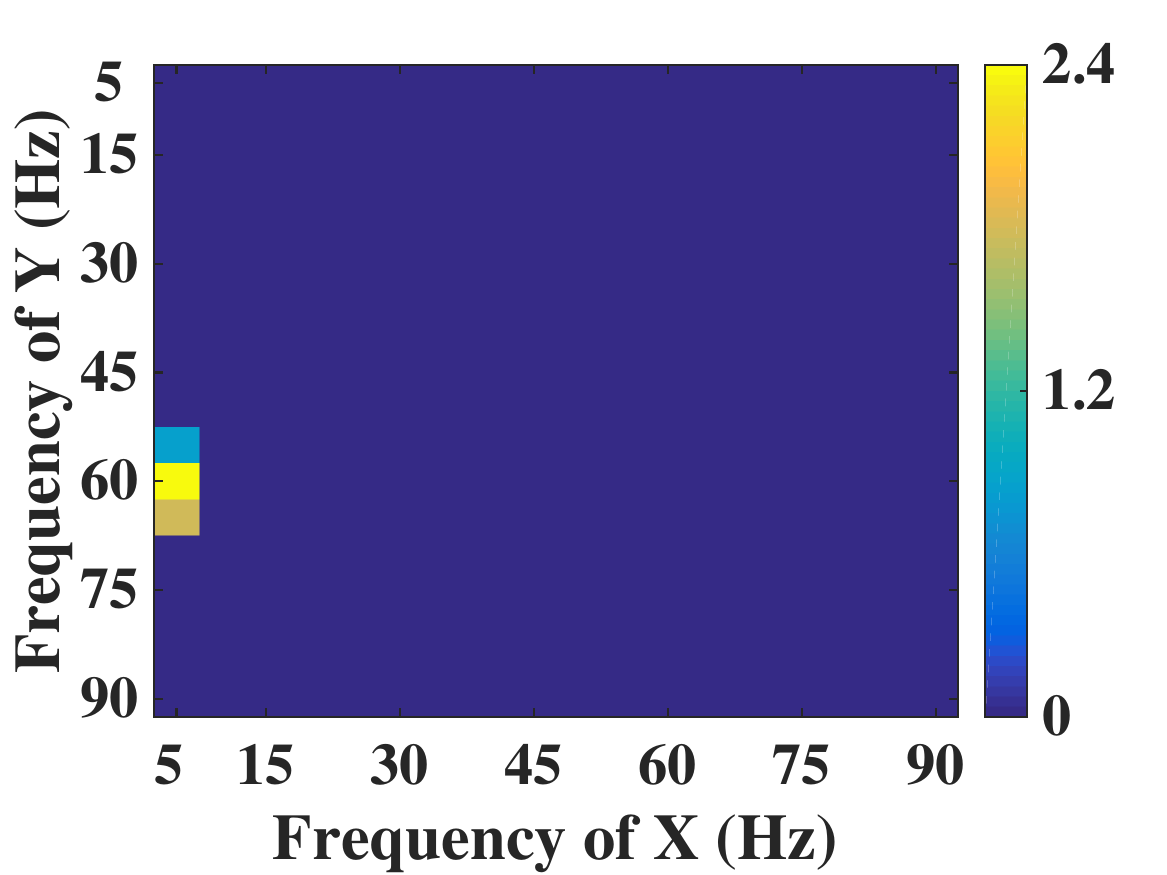} }
\hfill
\subfloat[Modulation Index]{
\includegraphics[width=.23\textwidth]{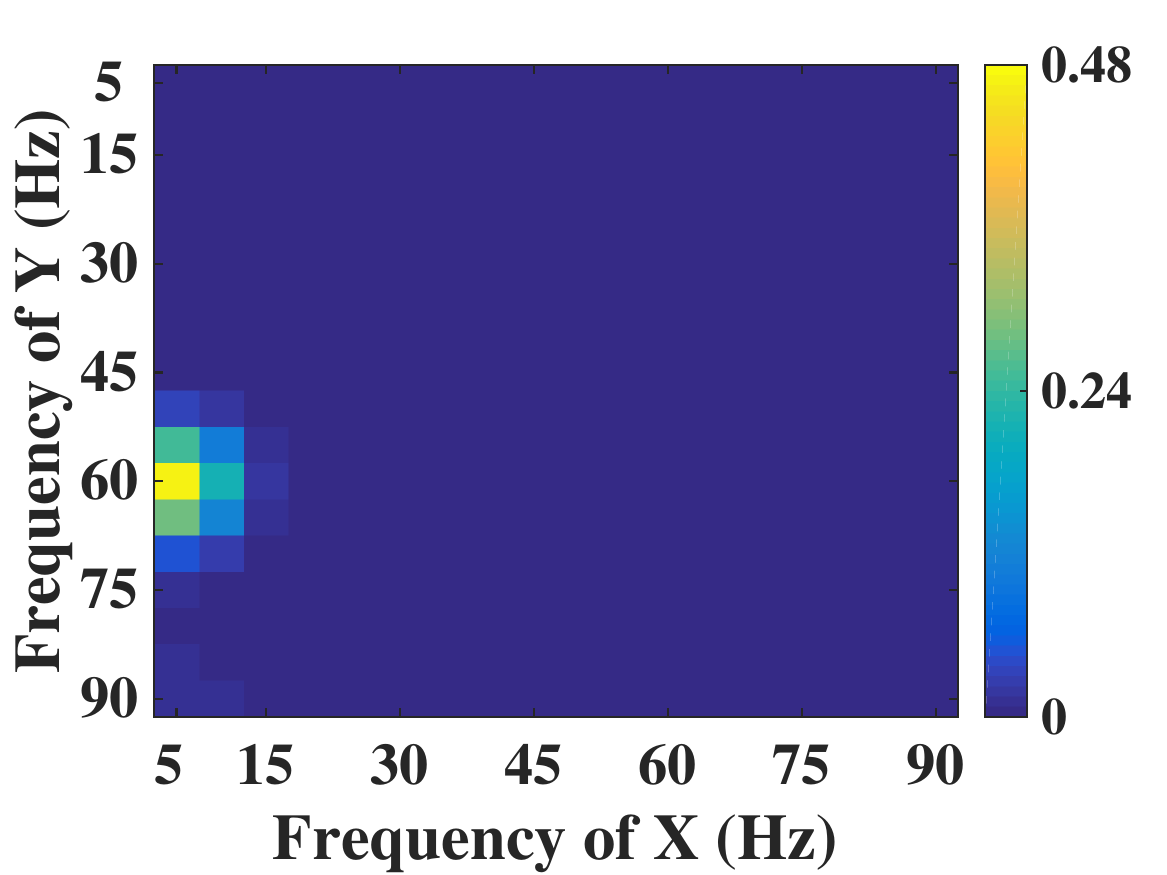} }
\hfill
\subfloat[MI-in-frequency]{
\includegraphics[width=.23\textwidth]{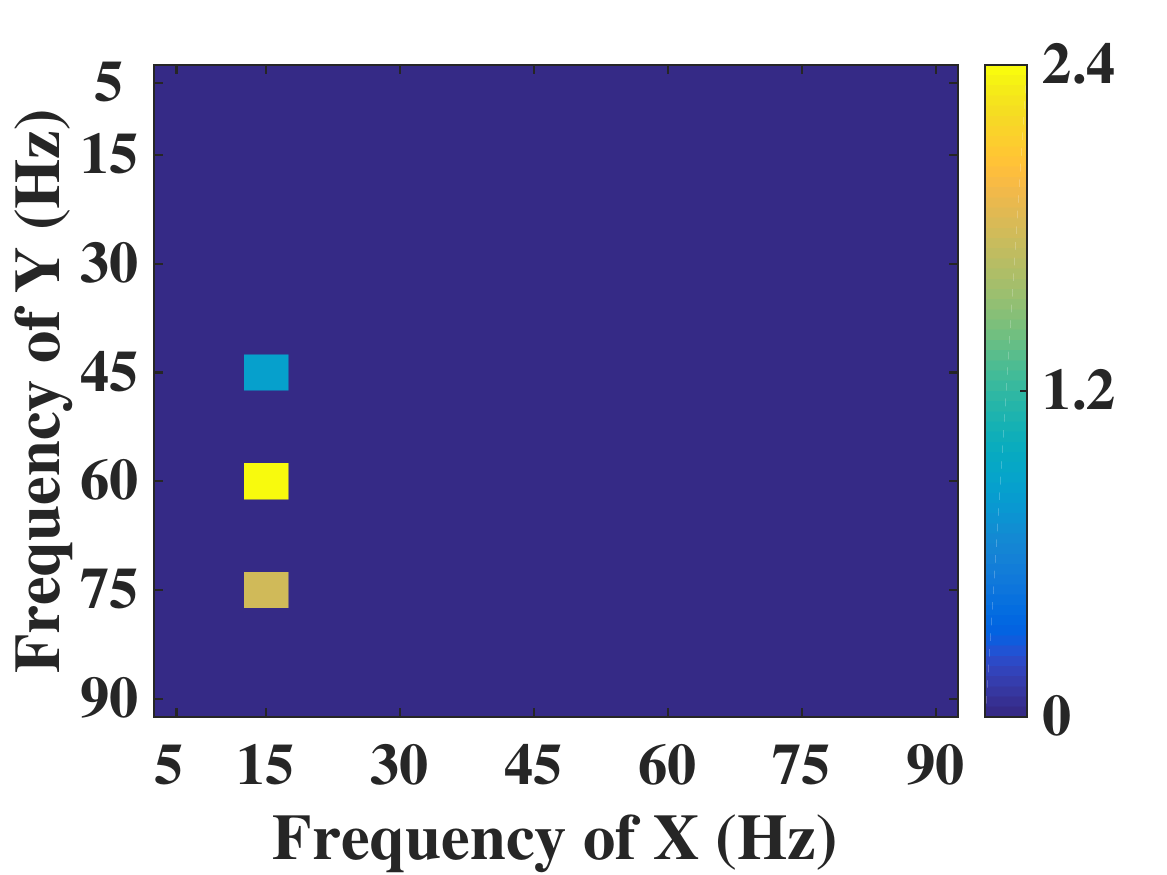} }
\hfill
\subfloat[Modulation Index]{
\includegraphics[width=.23\textwidth]{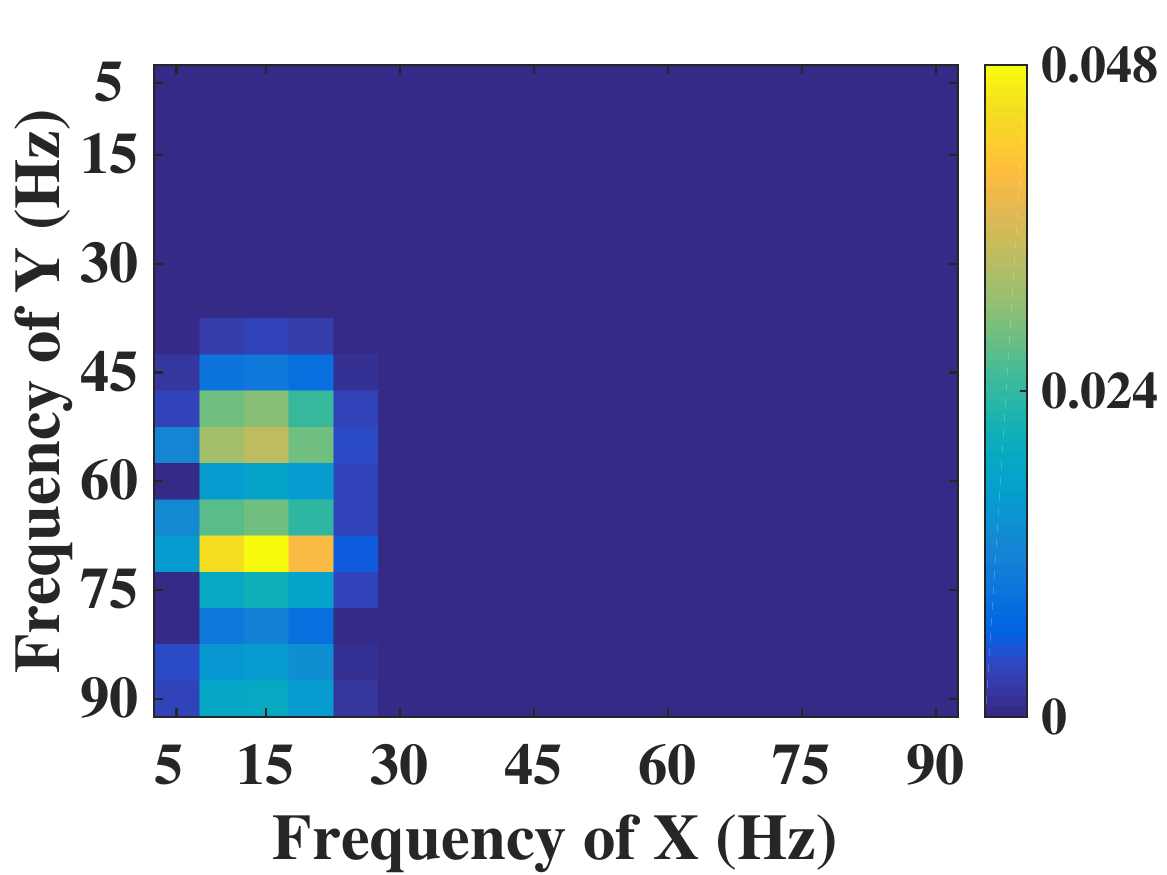} }
\caption{Comparing the performance of MI-in-frequency against modulation index in detecting cross-frequency coupling in data generated from \eqref{amp_modulation_eq}. In Fig.~\ref{Fig:amp_modulation}a and Fig.~\ref{Fig:amp_modulation}b, MI-in-frequency estimates obtained from nearest neighbor algorithm and modulation index are plotted respectively, when $f_l=5$ Hz and $f_h=60$ Hz in \eqref{amp_modulation_eq}. Fig.~\ref{Fig:amp_modulation}c and Fig.~\ref{Fig:amp_modulation}d respectively plot the MI-in-frequency estimates and modulation index estimates, when $f_l=15$ Hz and $f_h=60$ Hz in \eqref{amp_modulation_eq}.}\label{Fig:amp_modulation}
\squeezeup
\end{figure*}

\begin{figure*}[!t]
\centering
\subfloat[$\widehat{\MI}_{YY}\left( \lambda_i, \lambda_j\right)$]{
\includegraphics[width=.31\textwidth]{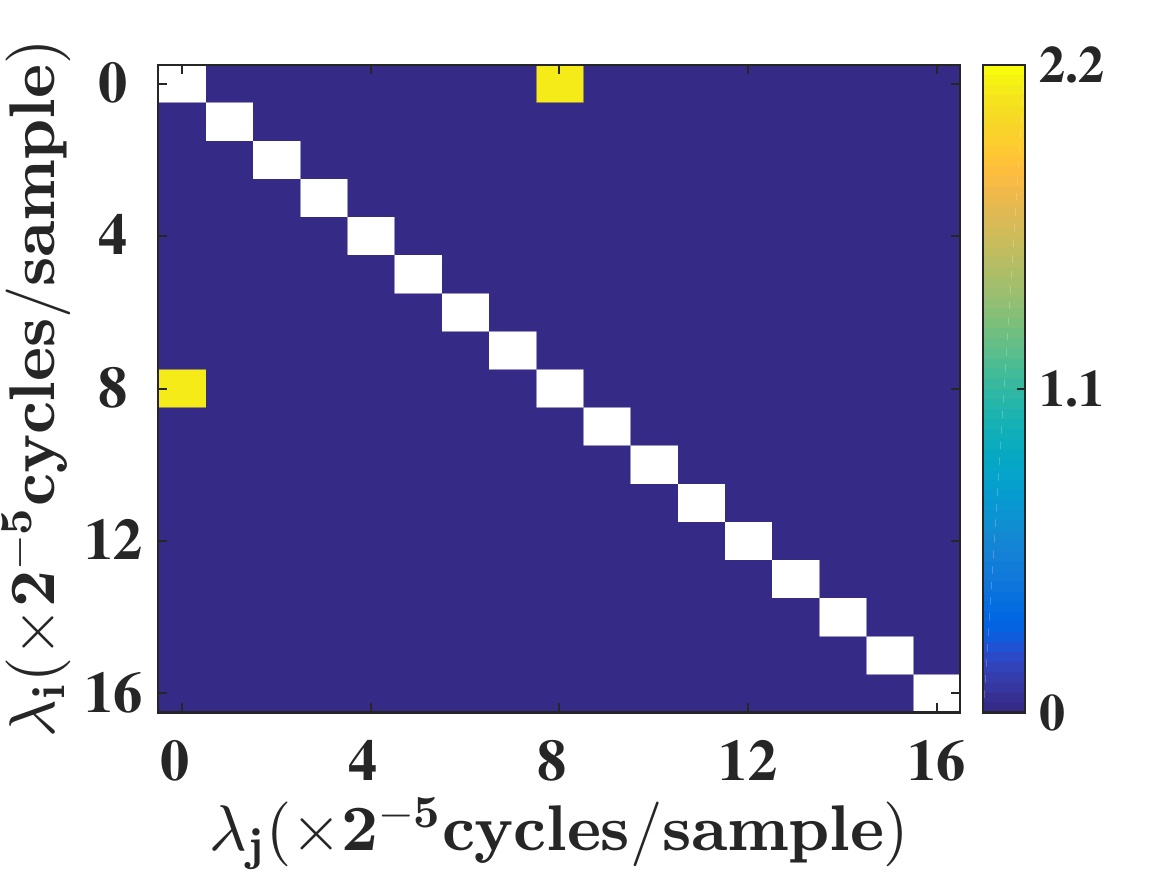}}
\hfill
\subfloat[$\widehat{\MI}_{XY}\left( \lambda_i, \lambda_j\right)$]{
\includegraphics[width=.31\textwidth]{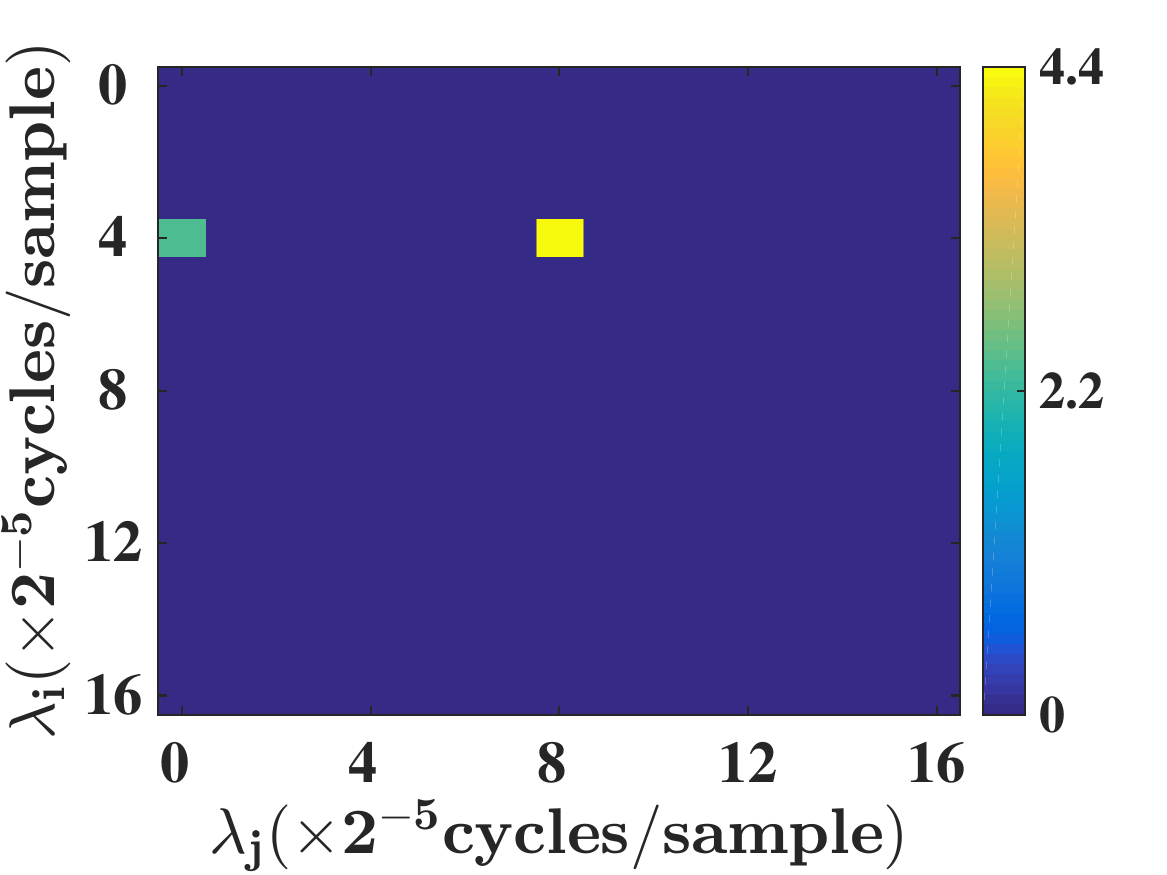}}
\hfill
\subfloat[MI between $X$ and $Y$]{
\includegraphics[width=.31\textwidth]{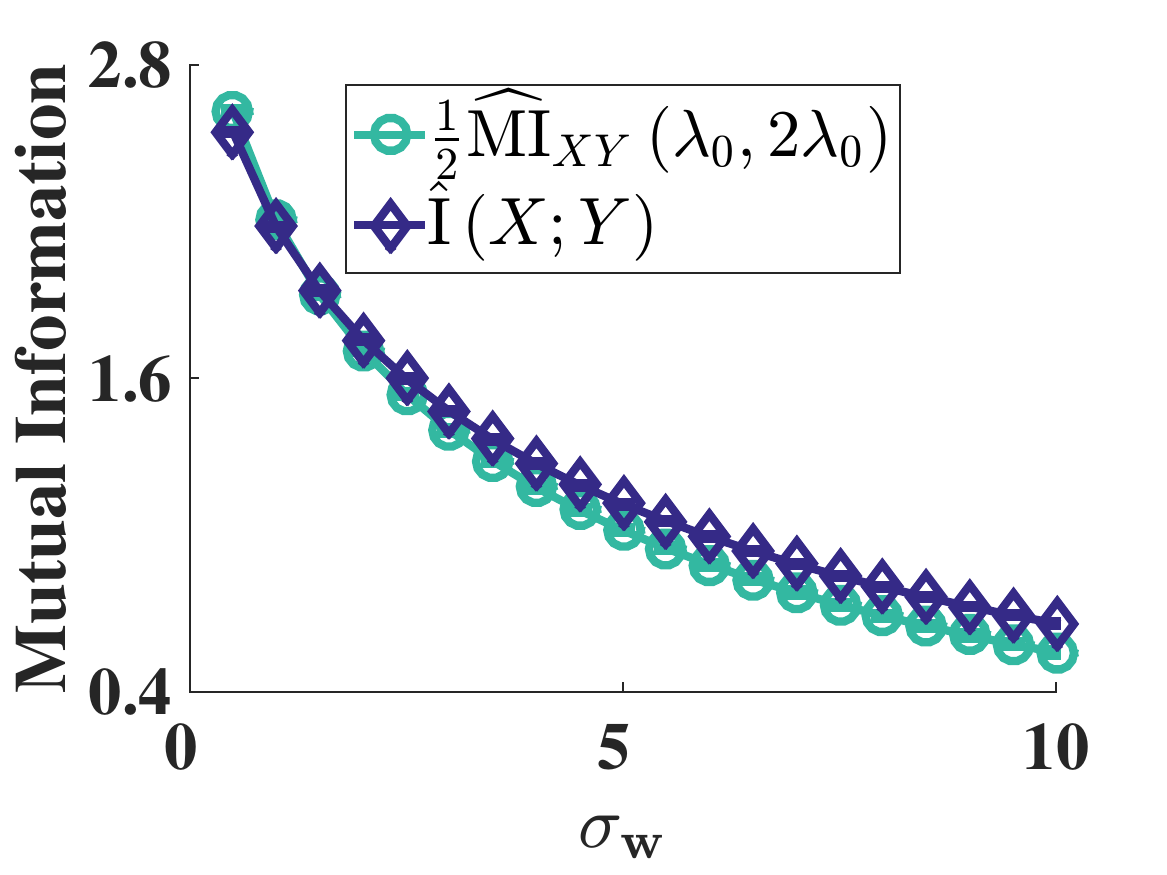} }
\caption{(a) MI-in-frequency estimates from the nearest neighbor based algorithm between the frequency components within the random processes $Y$, obtained from the single cosine data-generation model, \eqref{eq:one_cosine} with $\sigma_w = 1$. Note that the MI-in-frequency estimates along the principal diagonal are not plotted, since they are equal to $\infty$. (b) MI-in-frequency estimates between random processes $X$ and $Y$ related by the single cosine data-generation model with $\sigma_w = 1$. It is clear that MI-in-frequency estimator correctly identifies the pairwise frequency dependencies. (c) MI-in-frequency between $X$ at $\lambda_{0}$ and $Y$ at $2\lambda_{0}$, $\widehat{\MI}_{XY}\left( \lambda_{0}, 2\lambda_{0}\right)$, obtained from \eqref{eq_knn_est} along with the MI estimate between $X$ and $Y$, $\hat{\I}\left(X;Y\right)$, obtained from Algorithm~\ref{mi_est_algo} for various values of the noise standard deviation, $\sigma_w$.
}\label{Fig:cosine_square}
\squeezeup
\end{figure*}

\subsection{Comparison with Modulation Index} \label{subsec:comp_with_modulation_index}
We now compare the effectiveness of MI-in-frequency against modulation index in detecting cross-frequency coupling, using the simulated model commonly used to validate CFC metrics \cite{aru2015, onslow2011, berman2012}. Modulation index quantifies the relationship between the phase and amplitude envelopes extracted by the Hilbert transform \cite{canolty2006}. Consider two random cosine waves, $s_l[n]$ and $s_h[n]$, at frequencies $f_l$ and $f_h$ respectively. Let $f_s$ denote the sampling frequency. The samples of time-series $X$ and $Y$ are generated from the following model:
\begin{align} \label{amp_modulation_eq}
\!\!s_l[n] = A\cos\left(2\pi\frac{f_{l}}{f_s}n + \theta \right), \: & s_h[n] = A\cos\left(2\pi\frac{f_{h}}{f_s}n + \theta \right) \nonumber \\
x[n] = s_l[n] + w_1[n] , y[n] =& \left(1 + s_l[n]\right)s_h[n] + w_2[n],
\end{align}
where $A$ is a Rayleigh random variable with parameter $1$ and $\theta$ is a uniformly distributed random variable between $0$ and $2\pi$ that is independent of $A$. $w_1[n], w_2[n]$ are samples of i.i.d white Gaussian noise process with standard deviation $1$. We generated samples from this model with $f_l=5$ Hz, $f_h=60$ Hz and $f_s = 200$ Hz. MI-in-frequency between $X$ and $Y$ is estimated using the nearest neighbor based algorithm from $N = 40\times10^4$ samples with $N_s=10^4$ and plotted in Fig.~\ref{Fig:amp_modulation}a.  Modulation index between $X$ and $Y$ estimated by using the Matlab toolbox \cite{onslow2011}, with the amplitude envelope estimated by the Hilbert transform and is plotted in Fig.~\ref{Fig:amp_modulation}b. It is clear that both MI-in-frequency and modulation index successfully detect the cross-frequency coupling between $5$ Hz component of $X$ and $\left\{55, 60, 65 \right\}$ Hz components of $Y$ for these parameter values. We then generated $X$ and $Y$ from \eqref{amp_modulation_eq} with $f_l=15$ Hz and all other parameter values unchanged. Fig.~\ref{Fig:amp_modulation}c plots the MI-in-frequency estimates obtained via NNMIF algorithm and as expected, we detect the CFC between $15$ Hz component of $X$ and $\left\{45, 60, 75 \right\}$ Hz components of $Y$. However, modulation index, depicted in Fig.~\ref{Fig:amp_modulation}d, was not able to correctly detect the CFC between $X$ and $Y$ for these parameter values. In addition, the strength of the modulation index decreased from around $0.5$ when $f_l=5$ Hz in Fig.~\ref{Fig:amp_modulation}b to $0.05$ when $f_l=15$ Hz in Fig.~\ref{Fig:amp_modulation}d. This is because metrics like modulation index can only detect the CFC correctly with good frequency resolution only when one of the frequencies involved is very small compared to the other frequency. Otherwise, the bandwidth of the filter used to extract the phase and the amplitude envelope should be larger, which will reduce the frequency resolution in the estimated CFC (note the smearing in Fig.~\ref{Fig:amp_modulation}d, when compared to Fig.~\ref{Fig:amp_modulation}b) \cite{aru2015, berman2012}. In addition, we tested modulation index on data generated from \eqref{discrete_linear_eq} and \eqref{discrete_nonlinear_eq} and found that modulation index is unable to detect the cross-frequency coupling for these relationships. This is not surprising since the modulation index like metrics are tuned to detect CFC when the underlying coupling is of the form in \eqref{amp_modulation_eq}, whereas the MI-in-frequency defined in this paper overcomes this shortcoming, as evident from its performance on various simulated models.

 \begin{figure*}[!t]
\centering
\subfloat[$\widehat{\MI}_{YY}\left( \lambda_i, \lambda_j\right)$]{
\includegraphics[width=.32\textwidth]{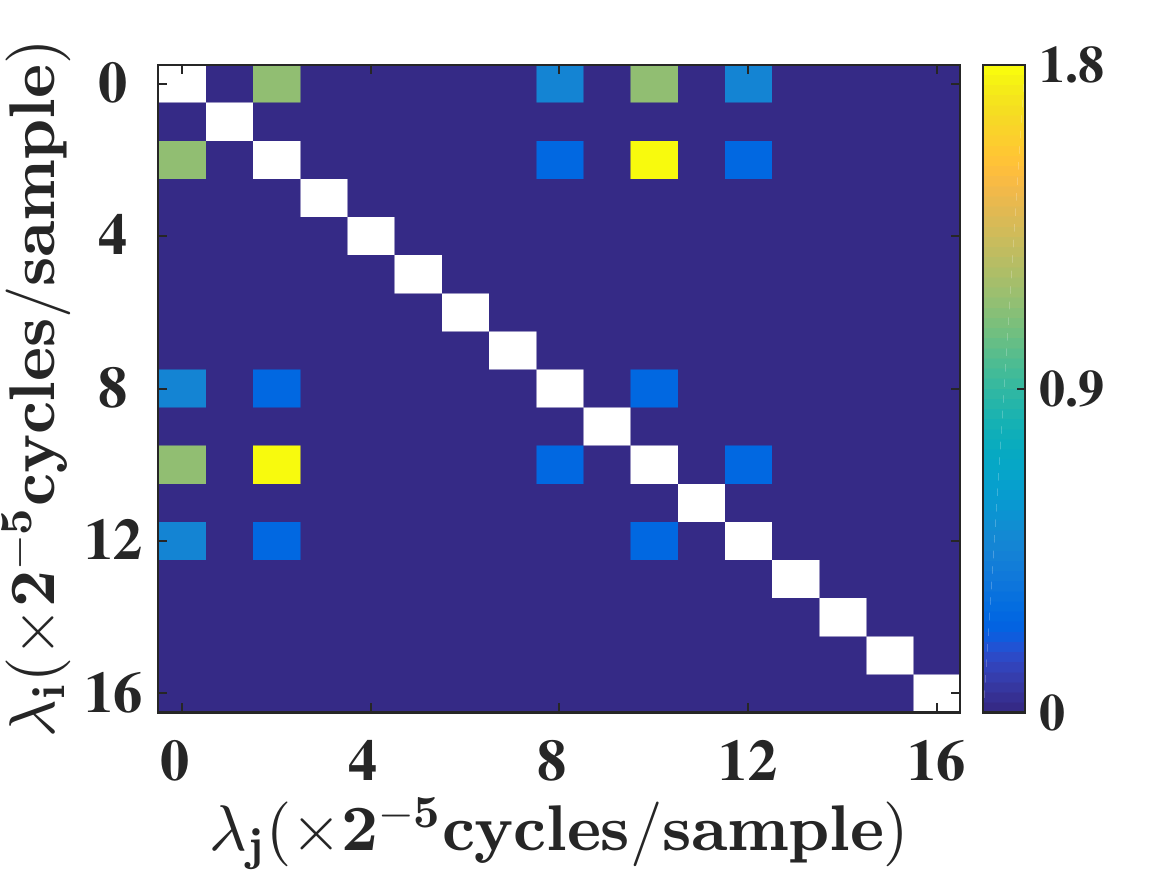}}
\hfill
\subfloat[$\widehat{\MI}_{XY}\left( \lambda_i, \lambda_j\right)$]{
\includegraphics[width=.32\textwidth]{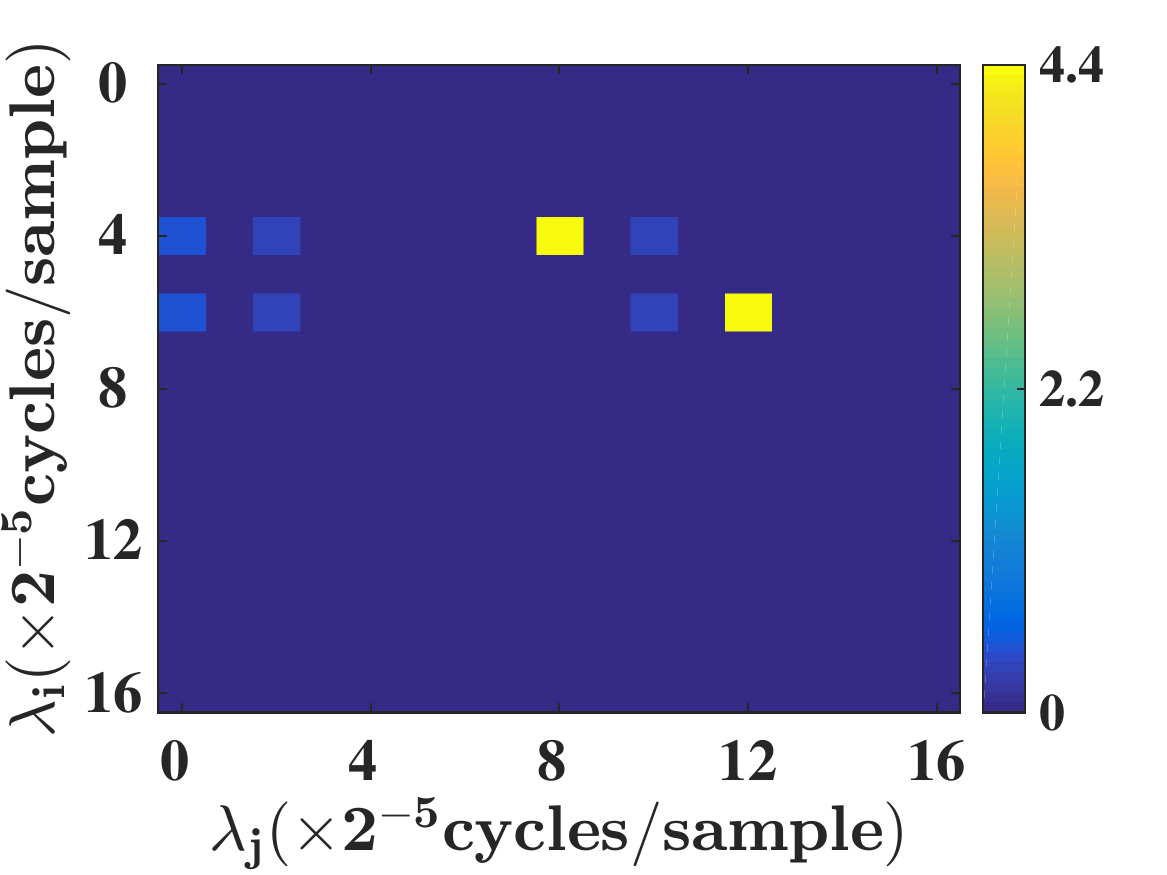}}
\hfill
\subfloat[MI between $X$ and $Y$]{
\includegraphics[width=.32\textwidth]{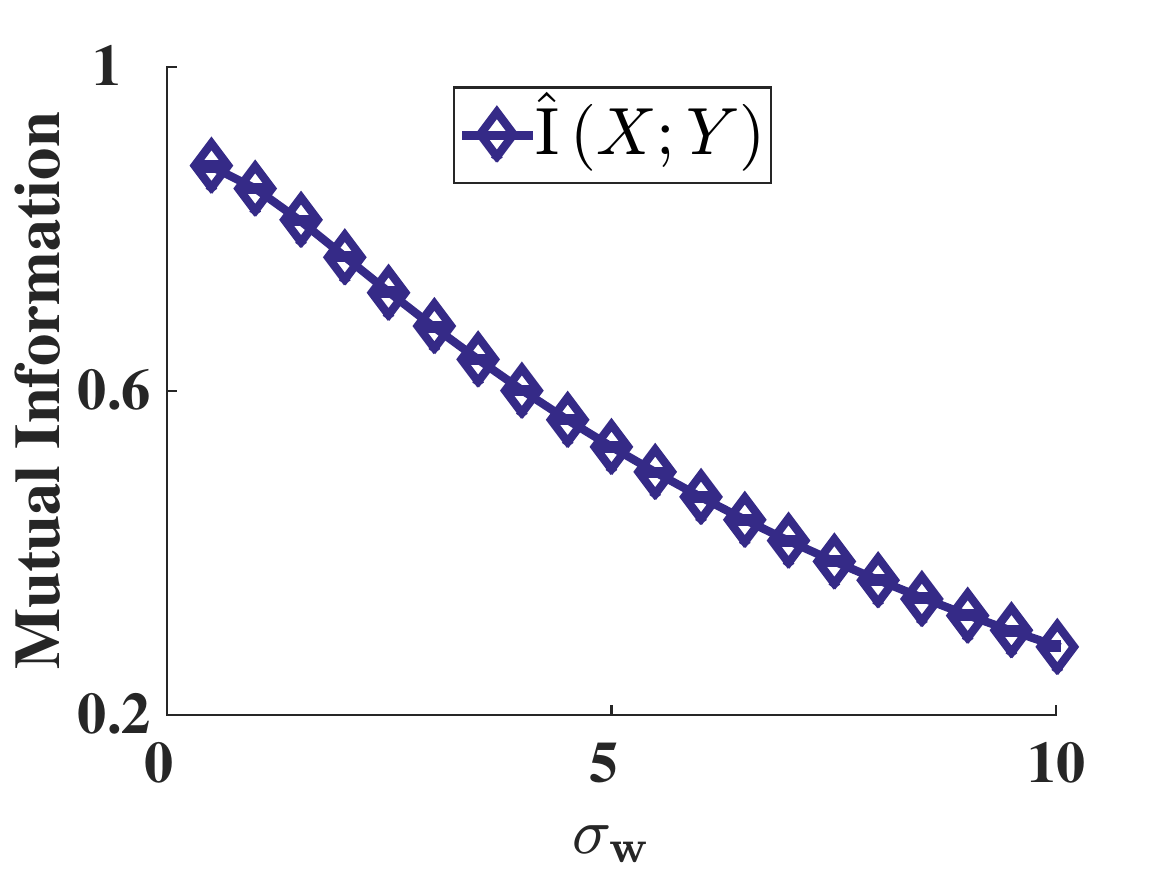} }
\caption{(a) MI-in-frequency estimates from the nearest neighbor based algorithm between the frequency components within the random processes $Y$, obtained from the two cosine data-generation model, \eqref{eq:two_cosines}. The MI-in-frequency estimates are not plotted along the diagonal, since they are equal to $\infty$. (b) MI-in-frequency estimates between random processes $X$ and $Y$ related by the two cosine data-generation model. It is clear that MI-in-frequency estimator correctly identifies the pairwise frequency dependencies between $X$ and $Y$. (c) $\hat{\I}\left(X;Y\right)$, the MI estimate between $X$ and $Y$ obtained from Algorithm~\ref{mi_est_algo} for various values of the noise standard deviation, $\sigma_w$.
}\label{Fig:two_cosine_square}
\squeezeup
\end{figure*}

\subsection{Nonlinear Models}
We now consider square nonlinearity, where the random processes $X$ and $Y$ are related by
\begin{align}\label{discrete_nonlinear_eq}
y[n] = x[n]^2 + w[n],
\end{align}
where $w[n]$ is white Gaussian noise with standard deviation $\sigma_w$. Modulation index was not able to detect and quantify the cross-frequency coupling for this model. We estimated the MI-in-frequency between frequency components within $Y$, $\widehat{\MI}_{YY}\left(\lambda_i, \lambda_j \right)$, between the frequency components of $X$ and $Y$, $\widehat{\MI}_{XY}\left(\lambda_i, \lambda_j \right)$, and the mutual information between $X$ and $Y$, $\hat{\I}\left(X;Y \right)$, from $N = 32\times10^4$ samples of $X$ and $Y$ with $N_s=10^4$, for different values of noise standard deviation, $\sigma_w \in \left[ 0, 10\right]$. Computing the true value of MI-in-frequency and mutual information is nontrivial because of the nonlinearity. The performance of the algorithms is assessed by checking if they detect the cross-frequency coupling at expected frequency pairs and by checking if the mutual information estimates decrease with increasing noise power as expected. We considered two different models for the stochastic process $X$, such that its samples are dependent across time.

\subsubsection{Random Cosine with Squared Nonlinearity} \label{sec:cosine_squared}
The samples of $X$ are generated from a random cosine wave, 
\begin{align} \label{eq:one_cosine}
x[n] = A\cos\left(2\pi\lambda_{0}n +\theta \right),
\end{align} 
where $A$ is a Rayleigh random variable with parameter $1$, $\theta$ is a uniform random variable between $0$ and $2\pi$ that is independent of $A$ and $\lambda_{0} = \frac{4}{32}$. It is easy to see that frequency components of $X$ are statistically independent and this is confirmed by the NNMIF estimator. However, because of the square nonlinearity in \eqref{discrete_nonlinear_eq}, the DC component of $Y$ and the $2\lambda_{0}$ component of $Y$ will be statistically dependent and this is confirmed by Fig.~\ref{Fig:cosine_square}a, which plots the MI-in-frequency between components of $Y$ generated with $\sigma_w=1$ using the NNMIF algorithm. The common information between these two processes will be present between $\lambda_{0}$ component of $X$ and the $\left\{0, 2\lambda_{0}\right\}$ components of $Y$. This cross-frequency dependence is confirmed by Fig.~\ref{Fig:cosine_square}b, which plots the estimates of MI-in-frequency between $X$ and $Y$ obtained by the NNMIF algorithm from \eqref{eq_knn_est}: we observe that significant  dependencies occur only at $\left(\lambda_{0}, 0 \right)$ and $\left(\lambda_{0}, 2\lambda_{0} \right)$ frequency pairs. As a result, $P=1, Q=2$.  The MI estimate from Algorithm~\ref{mi_est_algo}, $\hat{\I}\left(X;Y\right) =  \frac{1}{2}\hat{\I}\left(d\widetilde{X}(\lambda_{0}); \big\{d\widetilde{Y}(0), d\widetilde{Y}(2\lambda_{0})\big\} \right)$ is plotted in Fig.~\ref{Fig:cosine_square}c. The MI estimate decreases with increasing $\sigma_w$ as expected. In addition, we note for this model that the DC component of $Y$ does not contain any extra information about $X$, given the $2\lambda_{0}$ component of $Y$. Therefore, we expect  $\frac{1}{2}\hat{\I}\left(d\widetilde{X}(\lambda_{0}); \big\{d\widetilde{Y}(0), d\widetilde{Y}(2\lambda_{0})\big\} \right) = \frac{1}{2} \widehat{\MI}_{XY}\left(\lambda_{0}; 2\lambda_{0} \right)$, a result verified in Fig.~\ref{Fig:cosine_square}c, since the two curves are very close. 

\subsubsection{Two Random Cosines with Squared Nonlinearity}\label{sec:two_cosine_squared}
The samples of random process $X$ are generated according to 
\begin{align} \label{eq:two_cosines}
x[n] = A_1\cos\left(2\pi\lambda_{1}n +\theta_1 \right) + A_2\cos\left(2\pi\lambda_{2}n +\theta_2 \right),
\end{align} 
where $A_1, A_2$ are independent Rayleigh random variables with parameter $1$, $\theta_1, \theta_2$ are independent uniformly distributed random variables between $0$ and $2\pi$ that are independent of $A_1$, $A_2$, and $\lambda_{1} = \frac{4}{32}, \lambda_{2}=\frac{6}{32}$. As before, the frequency components of $X$ are statistically independent. However, after some basic algebra, it is easy to see that the all possible pairs of frequency components of $Y$ in $\left\{0, \lambda_{2} - \lambda_{1},  2\lambda_{1}, \lambda_{2} + \lambda_{1}, 2\lambda_{2} \right\}$  are statistically dependent, except for $\left(2\lambda_{1}, 2\lambda_{2}\right)$ frequency pair, and we expect to see statistically significant MI-in-frequency estimates between these frequency components. This is confirmed by Fig.~\ref{Fig:two_cosine_square}a, which plots the MI-in-frequency estimates within $Y$, generated with $\sigma_w=1$ and obtained by the NNMIF algorithm. The pairwise frequency dependencies between $X$ and $Y$ occur at  $\left(\lambda_{1}, 0 \right)$, $\left(\lambda_{1}, \lambda_{2} - \lambda_{1} \right)$, $\left(\lambda_{1}, 2\lambda_{1} \right)$,  $\left(\lambda_{1}, \lambda_{2} + \lambda_{1} \right)$, $\left(\lambda_{2}, 0 \right)$, $\left(\lambda_{2}, \lambda_{2} - \lambda_{1} \right)$, $\left(\lambda_{2}, \lambda_{2} + \lambda_{1} \right)$ and $\left(\lambda_{2}, 2\lambda_{2} \right)$.  Fig.~\ref{Fig:two_cosine_square}b plots the estimates of pairwise MI-in-frequency between $X$ and $Y$ generated with $\sigma_w=1$ and obtained by the data-driven NNMIF algorithm using \eqref{eq_knn_est}. The algorithm correctly identifies all the dependent frequency pairs and $P=2, Q=5$. We then apply the algorithm described in section~\ref{sec:mi_est} and plot the estimates the MI for different values of noise standard deviation $\sigma_w$ in Fig.~\ref{Fig:two_cosine_square}c. Again, the MI decreases with increasing noise power, as expected. These different models validate the superiority of MI-in-frequency over other existing metrics to detect cross-frequency coupling and also demonstrate the performance and accuracy of the data-driven MI-in-frequency and MI estimators.

\section{CFC in Seizure Onset Zone}\label{sec:cfc_in_soz}
\begin{table*}[!h]
\begin{center}
\caption{Clinical Details of the Patients Analyzed.}
\label{Table0}
\setlength{\tabcolsep}{6pt}
\blue{\begin{tabular}{c c c c c}
\hline
\hline
Patient  & {\makecell{ Number of Seizures \\ Analyzed}} & Age/Sex & {\makecell{Seizure Onset Zone}} & {\makecell{ Outcome \\ of Surgery}} \\
\hline
P1 & 3 & 22/M  & {\makecell{RAH 1-2, RPH 4, \\ RAMY 2-3}} & Class IA \\
\hline
P2  & 3 & 61/M  & {\makecell{LAH 2-4, LPH 2}} & Class IIIA \\
\hline
P3 & 2 & 29/M  & {\makecell{PD 4, 5 \\ LF 28, LP 4}} & Class IA \\
\hline
P4 & 3 & 21/F  & {\makecell{MST 1, TP 1, HD 1}} & Class IA \\
\hline
P5 & 3 & 24/M  & {\makecell{LPH 5, 6, LPSM 8, LMH 5, \\ RMH 4, 5, RPSM 7}} & Class IB \\
\hline
P6 & 3 & 35/M  & {\makecell{AH 3-5, PH 4 \\ AMY 2, 3}} & Class IA \\
\hline
P7 & 3 & 26/M  & {\makecell{AH 1, 2, 5, PH 5 \\ TOP 3, 4}} & Class IIB \\
\hline
P8 & 3 &  41/M & {\makecell{LAH 5, LAMY 3}} & N/A \\
\hline
P9 & 2 & 18/F  & {\makecell{RAH 3-5, LPH 6, \\ RPH 5-7}} & Class IB \\
\hline
\end{tabular}}
\end{center}
\begin{tablenotes}
\item \small{\blue{The full forms of the electrodes in seizure onset zone column in Table~\ref{Table0}: RAH - right anterior hippocampus, RPH - right posterior hippocampus, RAMY - right amygdala, PD - posterior hippocampal depth, MST - mid-subtemporal lobe, TP - temporopolar, HD - hippocampal depth and AST - anterior sub-temporal lobe, LMH - left mid hippocampus, AH - anterior hippocampus, PH - posterior hippocampus, AMY - amygdala, TOP - temporo-occipito-parietal. The outcomes are in Engel epilepsy surgery outcome scale \cite{engel1993, tonini2004}: ``Class IA - completely seizure free since surgery, class IB - non disabling simple partial seizures only since surgery, class IIB - rare disabling seizures since surgery (`almost seizure-free'), class IIIA - worthwhile seizure reduction, class IV - no worthwhile improvement".}}
\end{tablenotes}
\end{table*}

Epilepsy is a common neurological disorder characterized by repeated, unprovoked seizures. \blue{The seizure onset zone (SOZ) comprises regions of the brain that are responsible for generating and sustaining seizures \cite{luders2006}. Surgical resection of the seizure onset zone is the prescribed treatment for a large portion of medically refractory epilepsy patients with focal epilepsy. However, surgical resection risks damage to critical functional zones that are frequently adjacent or even overlapping with the seizure focus, depending on location of the focus \cite{gleissner2002}. An ideal solution might be a closed-loop neuromodulation strategy that stimulates the epileptic \cite{malladi2016, karunakaran2017} and other networks \cite{kim2016} at the optimal frequency with spatial and temporal specificity \cite{sunderam2010, krook2015}. In this paper, we focus on learning more about the characteristic frequencies and the spatial specificity of epileptic networks. Specifically, we investigate cross-frequency coupling between various regions in the seizure onset zone during the evolution of seizures and identify the frequencies with strong coupling. We estimate the cross-frequency coupling (CFC) from ECoG data recorded from the SOZ electrodes using our nearest neighbor based MI-in-frequency estimator. We infer the characteristics of CFC within and between various regions inside the seizure onset zone.}

\blue{We analyzed ECoG data, sampled at $F_s = 1$ kHz, from a total of 25 seizures recorded from nine patients with medial temporal lobe epilepsy. Clinical details of the patients, along with the seizure onset zone identified from ECoG data  \cite{malladi2016}, are summarized in Table~\ref{Table0}. The seizure start and end time were marked by the neurologist.} We analyzed  ECoG recordings from SOZ electrodes during preictal (window spanning up to 3 minutes immediately before the seizure starts), ictal (during seizures) and postictal (window spanning up to 3 minutes immediately after the seizure ends) periods. We only focussed on the oscillations in alpha (7.5-12.5 Hz), beta (12.5 - 30 Hz), gamma (30-80 Hz) and ripples (80-200 Hz), excluding $60$ Hz line noise and its harmonics. The CFC oscillations are analyzed at spectral resolution of $10$ Hz by choosing $N_f=100$, and the exact frequencies considered are $\{10,20,\cdots,200 \}$ Hz, excluding $\{60, 120, 180 \}$ Hz. The resulting $17 \times 17$ CFC matrix from each ECoG electrode and between all pairs of ECoG electrodes in the SOZ is estimated using nearest neighbor based estimator (section~\ref{sec:NNMIF}) during preictal, ictal and postictal periods during all the \blue{twenty five seizures}.

\begin{figure}[!t]
\centering
\includegraphics[width=0.7\columnwidth]{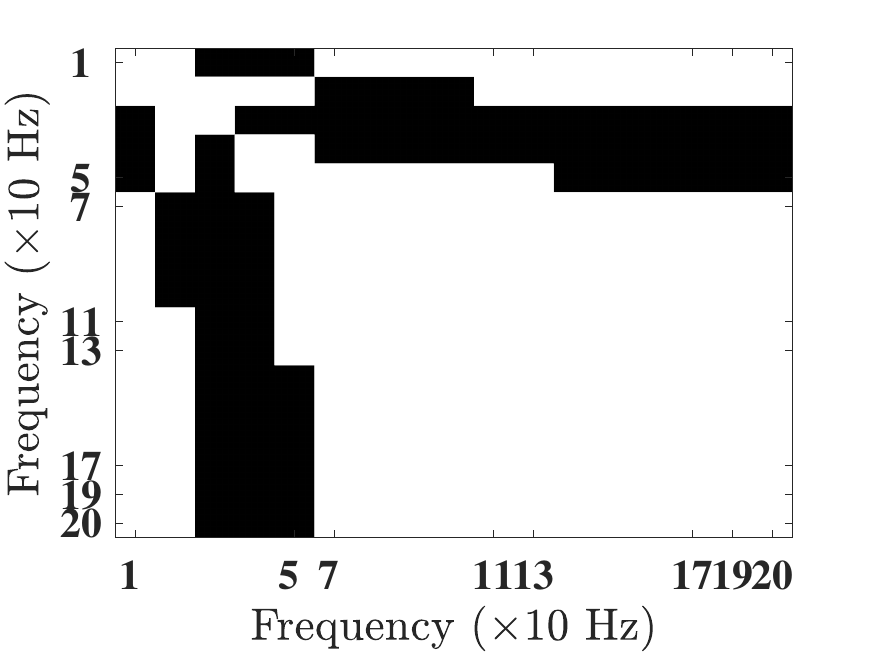}
\caption{\blue{Binary mask plotting the frequency pairs with statistically significant differences across ahypotheses tested after applying false discovery rate correction. White and black colored regions represent frequency pairs with and without statistically significant variation respectively.}}\label{Fig:binary_mask}
\squeezeup
\end{figure} 

\begin{figure*}[!t]
\centering
\subfloat[Preictal period]{
\includegraphics[width=0.31\textwidth]{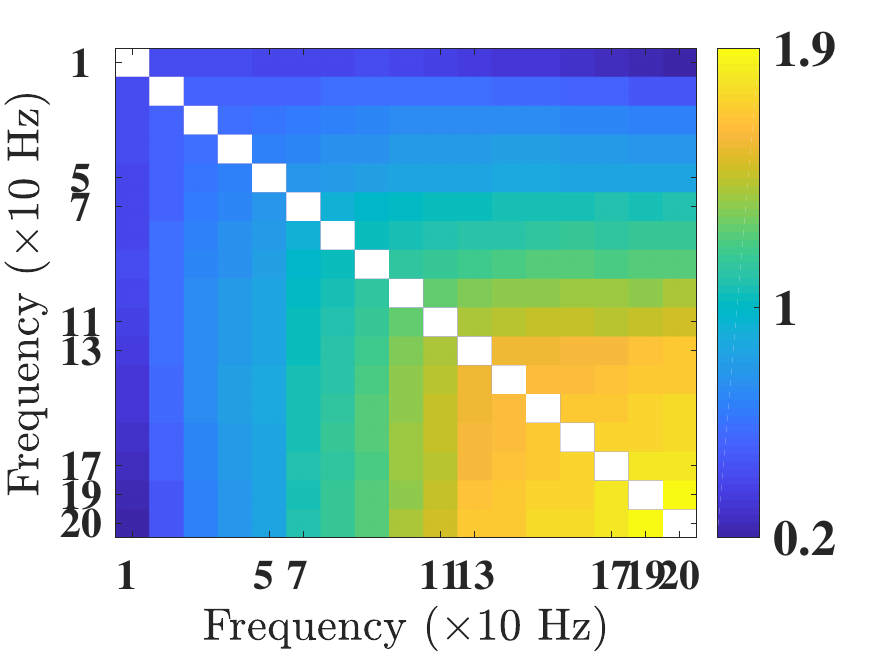} }
\hfill
\subfloat[Difference between ictal and preictal period]{
\includegraphics[width=0.31\textwidth]{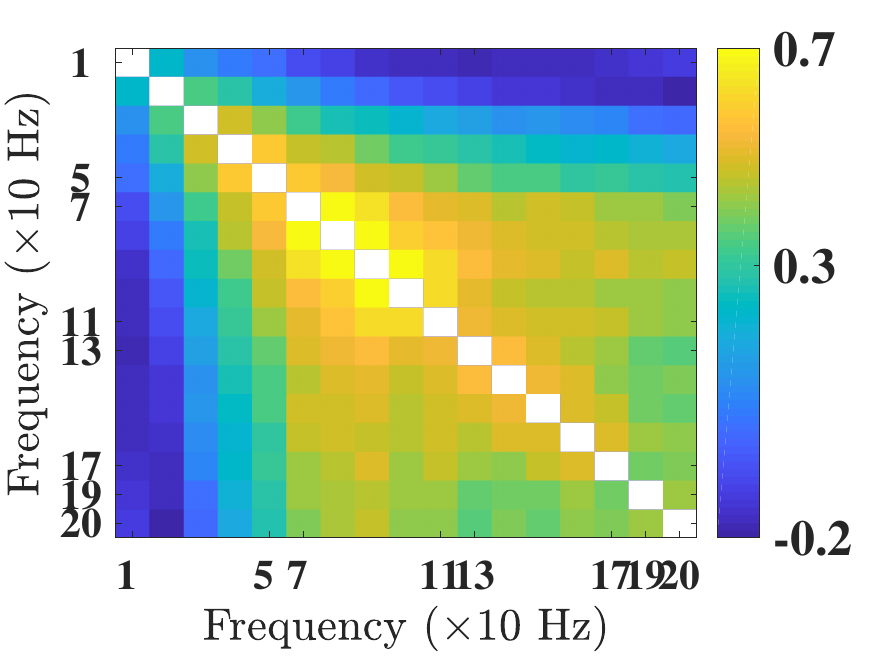} }
\hfill
\subfloat[Difference between postictal and ictal period ]{
\includegraphics[width=0.31\textwidth]{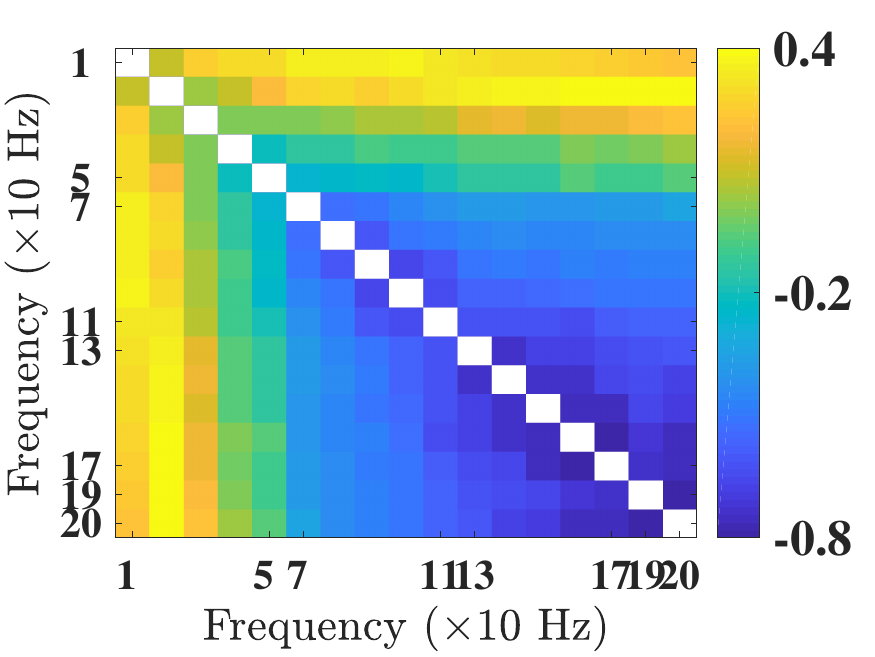} }
\caption{\blue{Cross-frequency coupling within an electrode inside the seizure onset zone. In Fig.~\ref{Fig:results_within}a, MI-in-frequency estimates over the frequencies $\left\{10,20,\cdots,200\right\}$ Hz excluding $\left\{60, 120, 180\right\}$ Hz are obtained from each electrode in SOZ during preictal period  and the median of the resulting CFC estimates from all the SOZ electrodes in the twenty five seizures from the nine temporal lobe epilepsy patients analyzed is plotted. In Fig.~\ref{Fig:results_within}b, MI-in-frequency estimates are obtained from each electrode in SOZ in the ictal period and the difference between the median CFC estimate in ictal and preictal period is plotted. Similarly, Fig.~\ref{Fig:results_within}c plots the difference in the median CFC between postictal and ictal periods.}}\label{Fig:results_within}
\end{figure*} 

\begin{figure*}[!t]
\centering
\subfloat[Preictal period]{
\includegraphics[width=0.31\textwidth]{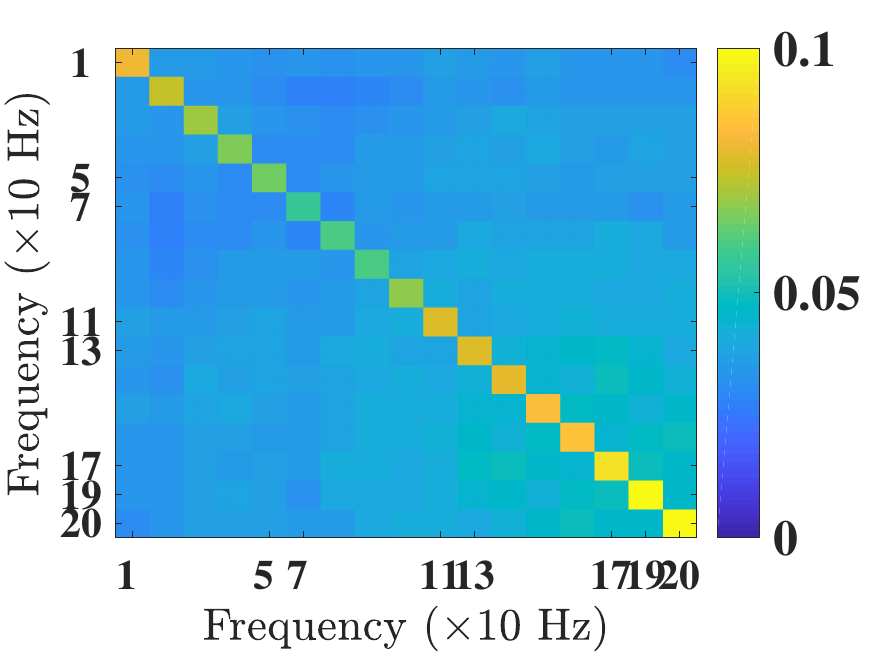} }
\hfill
\subfloat[Difference between ictal and preictal period]{
\includegraphics[width=0.31\textwidth]{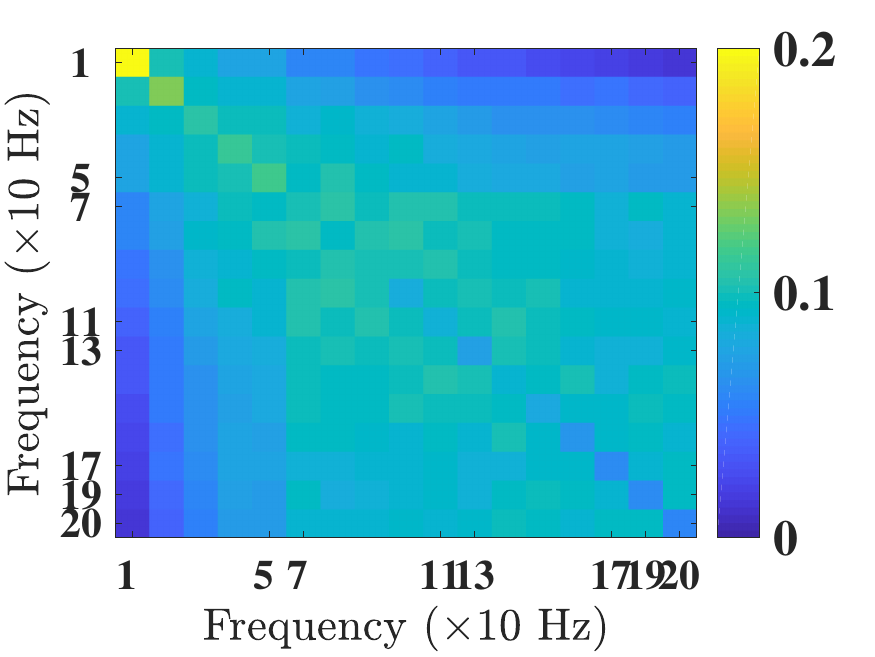} }
\hfill
\subfloat[Difference between postictal and ictal period]{
\includegraphics[width=0.31\textwidth]{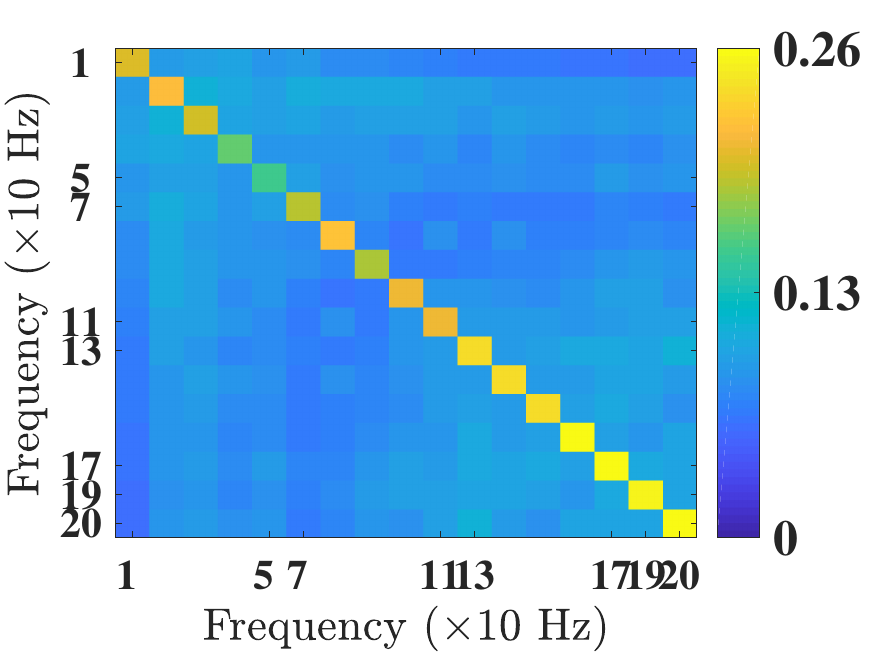} }
\caption{\blue{Cross-frequency coupling between electrodes in different regions inside the seizure onset zone. In Fig.~\ref{Fig:results_between}a, MI-in-frequency estimates over the frequencies $\left\{10,20,\cdots,200\right\}$ Hz excluding $\left\{60, 120, 180\right\}$ Hz are obtained between electrodes in different SOZ regions during the preictal period and the median of the resulting CFC estimates from the twenty five seizures in the nine temporal lobe epilepsy patients analyzed is plotted. In Fig.~\ref{Fig:results_between}b, MI-in-frequency estimates are obtained between electrodes in different SOZ regions from the ictal period and the difference between the median CFC estimate from the ictal and preictal period is plotted. Similarly, Fig.~\ref{Fig:results_between}c plots the difference in the median CFC between postictal and ictal periods.}}\label{Fig:results_between}
\end{figure*} 

We then grouped the ECoG electrodes into distinct anatomical regions based on their label and analyzed the average CFC within a SOZ electrode, between two electrodes in the same anatomical region and between electrodes in different anatomical regions. For instance, consider patient P1. \blue{ ECoG electrodes in the SOZ of patient P1 are grouped into three different anatomical regions--RAH, RPH, and RAMY (Table~\ref{Table0}). We estimated 5 CFC matrices, one per SOZ electrode, to infer the average CFC within an electrode in SOZ in this patient. We estimated 20 CFC matrices between all pairs of electrodes in the SOZ. Of these, 4 CFC matrices (2 to learn the CFC between the 2 SOZ electrodes in RAH and 2 to learn the CFC between the 2 SOZ electrodes in RAMY regions) are grouped to learn the average CFC between electrodes in the same anatomical region in SOZ. The remaining 16 CFC matrices are grouped to learn the CFC between different regions inside the SOZ. The estimated CFC matrices are grouped into these three spatial categories for all the nine patients during preictal, ictal and postictal periods. We only presented the results for CFC within a SOZ electrode and between electrodes in different SOZ regions during preictal, ictal and postictal periods.}

We used the permutation procedure outlined in section~\ref{sec:stat_sig} to estimate the CFC under the null hypothesis and assess the significance of the estimated CFC values across the six conditions considered (CFC during preictal, ictal, postictal periods within a SOZ electrode and between electrodes in different SOZ regions) using Wilcoxon signed-rank test \cite{corder2014}. We also used the Wilcoxon signed-rank test to identify the frequency pairs with significant variation in CFC between preictal and ictal periods and between ictal and postictal periods, both within a SOZ electrode and between electrodes in different SOZ regions (four hypotheses in total). In addition, we used the Mann-Whitney U-test \cite{corder2014} to identify frequency pairs with significant changes in CFC within a SOZ electrode and between electrodes in different SOZ regions across preictal, ictal and postictal periods (three hypotheses in total). We estimated $3621$ p-values in total ($13\times 17\times16 + 5\times17$) and applied false discovery rate correction at a significance level of $0.01$ to account for multiple comparisons \cite{benjamini1995}. The frequency pairs with significant statistical variation across all the hypotheses considered are depicted using a binary mask in Fig.~\ref{Fig:binary_mask}, in which black and white colored regions respectively represent frequency pairs without statistically significant variation and with statistically significant variation. Lack of statistical significance at the black regions in Fig.~\ref{Fig:binary_mask} could be because of insufficient data or could be due to a neuronal transition mechanism as the brain moves from preictal to ictal to postictal state. It is important to note that if we tested only a subset of the thirteen hypotheses, then some of the frequency pairs in black colored regions in Fig.~\ref{Fig:binary_mask} could become statistically significant.

\blue{The median CFC within an electrode in SOZ during preictal, ictal and postictal periods grouped across all twenty five seizures in nine patients analyzed is plotted in Fig.~\ref{Fig:results_within}.} In Fig.~\ref{Fig:results_within}a, median CFC in the preictal period is plotted, while the difference between median CFC in the ictal and preictal period, and between postictal and ictal period is plotted in Fig.~\ref{Fig:results_within}b and Fig.~\ref{Fig:results_within}c respectively. We need to multiply the binary mask in Fig.~\ref{Fig:binary_mask} with the plots in Fig.\ref{Fig:results_within} to obtain frequency pairs with significant statistical variation. The $(i,j)^{\mathrm{th}}$ element in the matrix in Fig.~\ref{Fig:results_within}a is the median MI-in-frequency between the $10i$ and $10j$ Hz frequency components during preictal period across all SOZ electrodes in the \blue{twenty five} seizures analyzed. The principal diagonal in the three CFC matrices is not plotted since MI-in-frequency between same frequencies in a signal is infinity. It is clear from this figure that ripple frequencies are heavily synchronized during preictal stage within an electrode in SOZ. The synchronization between all frequency pairs, particularly in gamma and ripples, seemed to increase during the seizure when compared to just before the seizure. And finally, the synchronization \blue{between} high-frequency bands decreased, and low frequencies become more synchronized \blue{amongst themselves and with high-frequencies} in the postictal period compared to the ictal period within an electrode in SOZ.

\blue{The median CFC between electrodes in different SOZ regions grouped across all twenty five seizures in nine patients analyzed is plotted in Fig.~\ref{Fig:results_between}.} We need to multiply the binary mask in Fig.~\ref{Fig:binary_mask} with the plots in Fig.~\ref{Fig:results_between} to obtain frequency pairs with significant statistical variation. The median CFC during the preictal period is plotted in Fig.~\ref{Fig:results_between}a. It is clear from the principal diagonal that neighboring regions in SOZ have \blue{weak}  linear interactions (possibly due to their spatial proximity) just before a seizure starts. From Fig.~\ref{Fig:results_within}a and Fig.~\ref{Fig:results_between}a, it is clear that the CFC strength is much lower between regions when compared to within an electrode. From Fig.~\ref{Fig:results_between}b, we observe a small increase in CFC between regions as the brain transitions to seizure state. However, the increase is much smaller between regions when compared to the increase observed in Fig.~\ref{Fig:results_within}b, which suggests that different SOZ regions potentially drive the rest of the brain into a seizure state independently, which implies any non-surgical treatment should target these different regions simultaneously to disrupt the epileptic network. As the brain transitions to postictal state, we observe a sharp increase in linear coupling between electrodes in different SOZ regions, which suggests that postictal periods, unlike ictal periods, are characterized by an increase in linear interactions. These results highlight the role of gamma and ripple high-frequency oscillations (HFOs) during seizures and the dynamic reorganization of synchronization between neuronal oscillations inside the seizure onset zone during the course of a seizure. \blue{These results also suggest that multiple regions inside the seizure onset zone might have to be targeted simultaneously using neuromodulation techniques to control seizure activity.}

\section{Discussion and Conclusions}
\blue{Detecting and quantifying relationships between multiple data streams recorded from a physical system is of interest in many science and engineering disciplines. However, since the underlying model is often unknown and nonlinear, detecting and quantifying the relationships in data is very challenging in most real-world applications. Brownian distance covariance \cite{szekely2009}, maximal information coefficient \cite{reshef2011} are some of the recent works that attempt to overcome this challenge in the most general case. Furthermore, in neuroscience, we are also interested in decomposing the relationships in frequency domain and estimating cross-frequency coupling (CFC) from electrophysiological recordings. Motivated to understand nonlinear frequency coupling in electrophysiological recordings from the brain and inspired by \cite{brillinger2007}, we defined MI-in-frequency between stochastic processes that are not necessarily Gaussian and estimated it using data-driven estimators. We found that the nearest neighbor based MI-in-frequency estimator outperforms the kernel-based MI-in-frequency estimator. MI-in-frequency can be thought of as `coherence' for non-Gaussian signals. At a first glance, CFC could be estimated by first filtering the data into appropriate frequency bands and then applying the techniques in \cite{szekely2009, reshef2011, onslow2011}. However, \cite{aru2015} summarizes all the caveats and confounds in estimating CFC using this approach. In contrast, the MI-in-frequency metric estimates CFC without explicitly band-pass filtering the data into appropriate frequency bands.}

We then compared the performance of MI-in-frequency against modulation index \cite{canolty2006, onslow2011}, a popular CFC metric used to measure phase-amplitude coupling that involves bandpass filtering, on simulated data and observed that MI-in-frequency outperforms the existing metrics used to estimate CFC. The main advantages of the MI-in-frequency approach over existing methods to estimate CFC are that it detects statistical independence, detects dependencies across phase and amplitude jointly, applies to linear and nonlinear dependencies, and is not dependent on parameters like the filter bandwidth. Our approach will need more data when compared with coherence since MI-in-frequency detects both linear and nonlinear dependencies in frequency. From the simulation results on linear models, we need about $10^3$ samples to be within $10\%$ of the true value. For the ECoG data sampled at $1$ kHz and a desired spectral resolution of $10$ Hz, this implies the total number of data samples is of the order of $100$ seconds or a couple of minutes, which is roughly the size of preictal, ictal and postictal windows used in section~\ref{sec:cfc_in_soz}. In summary, we developed a metric to detect statistical independence in frequency which outperforms existing CFC metrics and for the first time, utilized frequency domain to estimate mutual information over time between dependent data.

The MI-in-frequency metric can be further extended along several directions and some of them are outlined here. We can move to wavelet based analysis to improve the fixed time-frequency resolution of our Fourier-based approach in future work. The assumption of data stationarity in observation window (also assumed by most CFC metrics) can be potentially relaxed by utilizing time-frequency distributions and developing heuristics to measure the dependencies across frequency. However, the inherent trade-off involved is that we are not guaranteed to detect statistical independence. It is also possible to define and estimate conditional MI-in-frequency to eliminate indirect coupling estimated between two signals because of a third signal which is coupled to both.

We then apply the MI-in-frequency estimators to infer the coupling between neuronal oscillations before, during and after seizures in the seizure onset zone. Spatially, we used the electrode labels to identify the different regions in the SOZ. This is just one possible way to analyze the spatial variation in CFC. Some of the other possible options include using the distance between electrodes or using the underlying neuronal cell types to split the electrodes into different regions in SOZ. Our MI-in-frequency metric provides a framework that can be utilized to learn the CFC characteristics for any desired spatial grouping. In addition, the frequency resolution of our estimated CFC was constant and wavelet transform, instead of Fourier transform, can be utilized to provide greater resolution at lower frequencies.

\begin{figure}[!t]
\centering
\subfloat[Within a SOZ electrode]{
\includegraphics[width=0.23\textwidth]{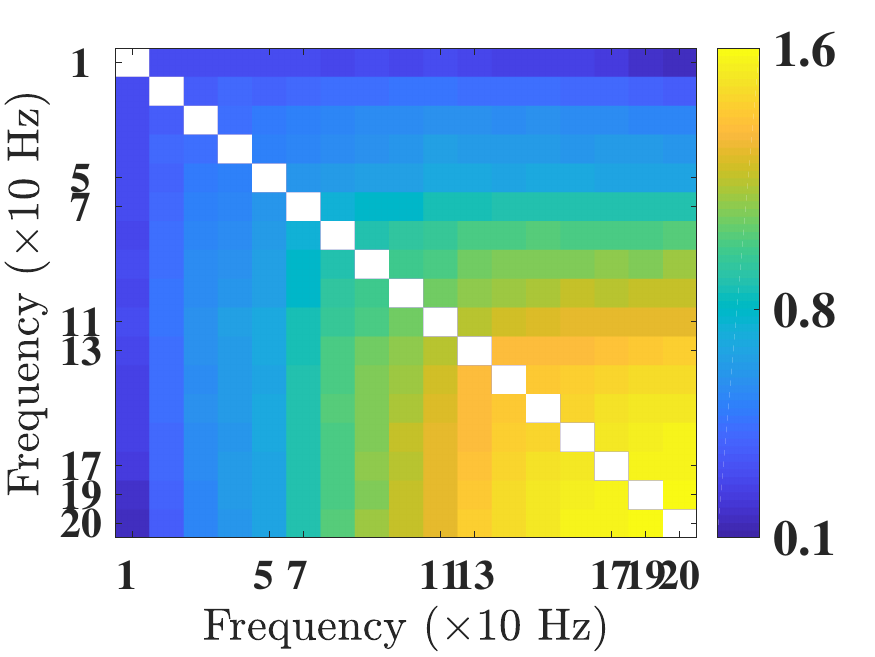} }
\hfill
\subfloat[Between electrodes in different SOZ regions]{
\includegraphics[width=0.23\textwidth]{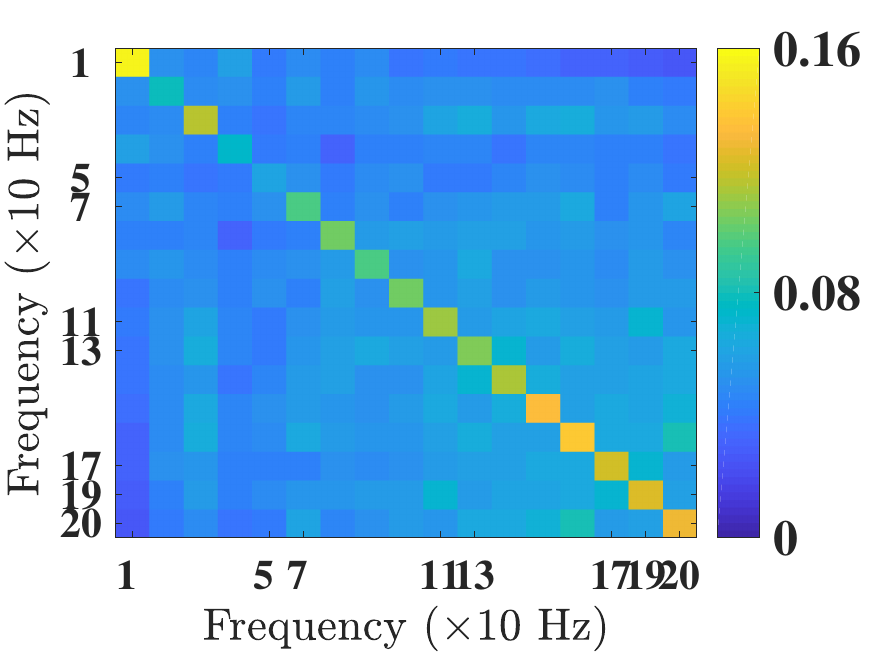} }
\caption{\blue{Cross-frequency coupling during interictal periods. In Fig.~\ref{Fig:results_interictal}a, MI-in-frequency estimates over the frequencies $\left\{10,20,\cdots,200\right\}$ Hz excluding $\left\{60, 120, 180\right\}$ Hz are obtained from each electrode in SOZ during interictal period  and the median of the resulting CFC estimates from all the SOZ electrodes in patients P1 and P2 is plotted. In Fig.~\ref{Fig:results_interictal}b, MI-in-frequency estimates over the frequencies $\left\{10,20,\cdots,200\right\}$ Hz excluding $\left\{60, 120, 180\right\}$ Hz are obtained between electrodes in different SOZ regions during the interictal period and the median of the resulting CFC estimates in patients P1 and P2 is plotted.}}\label{Fig:results_interictal}
\squeezeup
\end{figure} 

We observed that the high-frequency synchronization within an ECoG electrode in SOZ increases during seizures and decreases immediately after the seizure, which is accompanied by an increase in low-frequency coupling. However, the coupling between different anatomical regions in SOZ does not increase noticeably during seizures and is also followed by a large increase in linear interactions immediately after a seizure. These observations suggest that seizure activity is characterized by nonlinear interactions and is potentially due to the independent efforts by various regions within SOZ, which implies that all these regions are potential spatial targets for electrical stimulation. \blue{Furthermore, we did a preliminary investigation to learn if there are the differences in CFC between interictal periods and seizure periods. Fig.~\ref{Fig:results_interictal} plots the CFC within an ECoG electrode and between ECoG electrodes in different regions in SOZ during interictal period in two patients (P1 and P2). Comparing Fig.~\ref{Fig:results_interictal} with  Fig.~\ref{Fig:results_within}a and Fig.~\ref{Fig:results_between}a, it looks like the CFC within a SOZ electrodes at higher frequencies slightly increases, while CFC between electrodes in different regions across the diagonal (or equivalently,  linear interactions) slightly decreases as the brain transitions from interictal to preitctal periods. We plan to extend this analysis to a larger patient cohort. Building a real-time seizure prediction system utilizing the variations in CFC between interictal and seizure periods is the focus of our current \cite{hooper2017} and future work.}  In addition, the CFC characteristics were patient-specific and we presented the median CFC across all the patients considered. Going forward, the MI-in-frequency metric should be applied to infer the CFC between channels in SOZ and outside SOZ to learn how SOZ drives the rest of the brain into a seizure state in each epilepsy patient. The results from such an analysis will improve our understanding of the CFC mechanisms underlying seizure activity and will serve as the first step towards the development of a patient-specific, closed-loop, non-surgical treatment for epilepsy.

\section{Acknowledgments}
The authors wish to thank Suganya Karunakaran for the helpful discussions on statistical hypothesis testing and proofreading the manuscript.

\bibliographystyle{IEEEtran}
\bibliography{refs}

\section{Appendix} \label{sec:appendix}
\paragraph{Proof of \eqref{Y_spec_decomp}} We have from \eqref{mvar_cont_eq},
\begin{align} \label{eq1}
& \qquad y(t)  = \int\limits_{-\infty}^{\infty}\!\! h_1(t-\tau) x(\tau) d\tau \!\! + \!\! \int\limits_{-\infty}^{\infty}\!\! h_2(t-\tau) w(\tau) d\tau \\
&\Rightarrow  \int\limits_{-\infty}^{\infty} e^{j2\pi \nu t} d\widetilde{Y}\left(\nu \right) = \int\limits_{-\infty}^{\infty} h_1(t-\tau) \int\limits_{-\infty}^{\infty} e^{j2\pi \nu \tau} d\widetilde{X}\left(\nu \right) d\tau  \nonumber \\
&\qquad+ \int\limits_{-\infty}^{\infty} \!\!h_2(t-\tau) \int\limits_{-\infty}^{\infty} e^{j2\pi \nu \tau} d\widetilde{W}\left(\nu \right)  d\tau \: \text{(from Theorem~\ref{theorem1})} \nonumber \\
& = \!\! \int\limits_{-\infty}^{\infty}\!\! e^{j2\pi \nu t} \!\! \int\limits_{-\infty}^{\infty} \!\! h_1(t-\tau)e^{-j2\pi \nu (t-\tau)}d\tau d\widetilde{X}\left(\nu \right) \!\! + \nonumber \\
& \qquad \qquad \int\limits_{-\infty}^{\infty}\!\! e^{j2\pi \nu t} \!\! \int\limits_{-\infty}^{\infty}\!\! h_2(t-\tau)e^{-j2\pi \nu (t-\tau)}d\tau d\widetilde{W}\left(\nu \right) \\
& = \!\! \int\limits_{-\infty}^{\infty}\!\! e^{j2\pi \nu t} \left\{H_1\left(\nu\right)d\widetilde{X}\left(\nu \right) \!\!+ \!\!H_2\left(\nu\right)d\widetilde{W}\left(\nu \right)\right\}.\!\! \\
& \implies d\widetilde{Y}\left(\nu \right) = H_1\left(\nu\right)d\widetilde{X}\left(\nu \right) + H_2\left(\nu\right)d\widetilde{W}\left(\nu \right). \nonumber
\end{align}

\paragraph{Proof of Theorem~\ref{theorem3}} We will first prove that $\MI_{XY}\left(\nu_1,\nu_2\right)$ is zero, when $X$ and $Y$ are related by \eqref{mvar_cont_eq} and $\nu_1\neq \nu_2$. Since the processes $X\left(t\right)$ and $W\left(t\right)$ are independent, their spectral processes are also independent. In addition, we also know from Theorem~\ref{theorem2} that the spectral increments of the Gaussian process $X\left(t\right)$ are independent. It is clear from \eqref{Y_spec_decomp} that given $H_1\left(\nu\right)$ and $H_2\left(\nu\right)$, $\big[d\widetilde{Y}_R\left( \nu_2\right),d\widetilde{Y}_I\left( \nu_2\right) \big]$ is completely determined by the two-dimensional random vectors $\big[d\widetilde{X}_R\left( \nu_2\right),d\widetilde{X}_I\left( \nu_2\right) \big]$ and $\big[d\widetilde{W}_R\left( \nu_2\right),d\widetilde{W}_I\left( \nu_2\right) \big]$, both of which are independent of the two-dimensional random vector $\big[d\widetilde{X}_R\left( \nu_1\right),d\widetilde{X}_I\left( \nu_1\right) \big]$ when $\nu_1 \neq \nu_2$. This implies the MI between $\big[d\widetilde{Y}_R\left( \nu_2\right),d\widetilde{Y}_I\left( \nu_2\right) \big]$ and $\big[d\widetilde{X}_R\left( \nu_1\right),d\widetilde{X}_I\left( \nu_1\right) \big]$, which is defined as $\MI_{XY}\left(\nu_1,\nu_2\right)$, is zero. 

We will now derive the analytical expression for $\MI_{XY}\!\left(\!\nu\!,\!\nu\!\right)$, for $\nu \neq 0$. Let $H_1\!\left(\nu\right)\!\! = \!\! H_{1R}\!\left(\nu\right) \!\! + \!\! j H_{1I}\!\left(\nu\right)$ and $H_2\!\left(\nu\right) \!\! = \!\! H_{2R}\!\left(\nu\right) \!\! + \!\! j H_{2I}\!\left(\nu\right)$. We can see from \eqref{XW_spec_decomp}, \eqref{Y_spec_decomp} that
\begin{align} \label{Y_spec_inc_dist}
 \big[d\widetilde{Y}_R\left(\nu\right),d\widetilde{Y}_I\left(\nu\right) \big]^{} \!\! \sim \!\! \mathcal{N} & \left(\mathbf{0},\left(\frac{1}{2}s_X\left(\nu\right)|H_1\left(\nu\right)|^2 \right. \right. + \nonumber \\
 & \left. \left. \frac{1}{2}s_W\left(\nu\right)|H_2\left(\nu\right)|^2 \right)\mathbf{I} \right),
\end{align}
where $\mathcal{N}$ represents Gaussian distribution, $\mathbf{0}$ is a two element zero vector and $\mathbf{I}$ is the $2 \times 2$ identity matrix. In addition, 
\begin{align} \label{XY_spec_inc_dist}
\hspace{-.2cm} \big[d\widetilde{X}_R\left(\nu\right)\!,\!d\widetilde{X}_I\left(\nu\right)\!,\!d\widetilde{Y}_R\left(\nu\right)\!,\!d\widetilde{Y}_I\left(\nu\right)\! \big]^{} \!\! \sim \!\! \mathcal{N}\!\!\left(\!\!\mathbf{0}, \! \begin{bmatrix}
					\Sigma_{11}\! &\! \Sigma_{12} \\
					\Sigma_{21}\! &\! \Sigma_{22} 
				\end{bmatrix} \!\right)\!\!,\!\!
\end{align}
where $\Sigma_{11} = \frac{1}{2}s_X\left(\nu\right) \mathbf{I}$, $\Sigma_{22} = \frac{1}{2} \sigma_{\widetilde{Y}}^2\left(\nu\right)\mathbf{I}$, $\sigma_{\widetilde{Y}}^2\left(\nu\right) = \left(s_X\left(\nu\right)|H_1\left(\nu\right)|^2+s_W\left(\nu\right)|H_2\left(\nu\right)|^2 \right)$, $\mathbf{I}$ is the $2 \times 2$ identity matrix and $\mathbf{0}$ is a four element zero vector. In addition, 
\begin{equation}
\Sigma_{12} = \Sigma_{21}^\mathrm{T} = \begin{bmatrix}
\frac{1}{2}s_X\left(\nu\right)H_{1R}\!\left(\nu\right) & \frac{1}{2}s_X\left(\nu\right)H_{1I}\!\left(\nu\right) \\
					-\frac{1}{2}s_X\left(\nu\right)H_{1I}\!\left(\nu\right) & \frac{1}{2}s_X\left(\nu\right)H_{1R}\!\left(\nu\right) 
				\end{bmatrix}. \nonumber
\end{equation}
Now, the MI between $X$ and $Y$ at frequency $\nu$ is given by
\begin{align}
& \MI_{XY}\left(\nu,\nu\right) = \I\big(\big\{d\widetilde{X}_R\left( \nu\right),d\widetilde{X}_I\left( \nu\right) \big\}; \big\{d\widetilde{Y}_R\left( \nu\right),d\widetilde{Y}_I\left( \nu\right) \big\} \big) \nonumber \\
& = \I\big(\big\{d\widetilde{X}_R\left( \nu\right),d\widetilde{X}_I\left( \nu\right) \big\}; d\widetilde{Y}_R\left( \nu\right)\big) + \nonumber \\
& \qquad \qquad \I\big(\big\{d\widetilde{X}_R\left( \nu\right),d\widetilde{X}_I\left( \nu\right) \big\}; d\widetilde{Y}_I\left( \nu\right)|d\widetilde{Y}_R\left( \nu\right) \big) \label{MI_chain_rule} 
\end{align}
\begin{align}
 = \I\big(\big\{d\widetilde{X}_R\left( \nu\right),&d\widetilde{X}_I\left( \nu\right) \big\}; d\widetilde{Y}_R\left( \nu\right)\big) + \nonumber \\
& \I\big(\big\{d\widetilde{X}_R\left( \nu\right),d\widetilde{X}_I\left( \nu\right) \big\}; d\widetilde{Y}_I\left( \nu\right)\big), 
\label{Spec_comp_independent}
\end{align}
where \eqref{MI_chain_rule} follows from the chain rule of mutual information \cite{cover2012} and \eqref{Spec_comp_independent} follows because the real and imaginary parts of the spectral process of a Gaussian process are independent from Theorem~\ref{theorem2}. In addition, $\big[d\widetilde{X}_R\left(\nu\right),d\widetilde{X}_I\left(\nu\right),d\widetilde{Y}_R\left(\nu\right) \big]^{}$ is a Gaussian distributed random vector with zero mean and covariance matrix $\Sigma^{\prime}$, which is easily obtained from \eqref{XY_spec_inc_dist}. Since the mutual information between components of a Gaussian random vector depends only on the determinants of the joint distribution's covariance matrices  and that of marginals \cite{cover2012}, we can easily show that 
\begin{align} \label{MI_3D_RIR}
 \I\big(\big\{d\widetilde{X}_R\left( \nu\right),d\widetilde{X}_I\left( \nu\right)& \big\}; d\widetilde{Y}_R\left( \nu\right)\big) =  \frac{1}{2} \log \frac{|\Sigma_{11}| \left(\frac{1}{2}\sigma_{\widetilde{Y}}^2\right)}{|\Sigma^{\prime}|} \nonumber \\
 & = \frac{1}{2}\log \big(1 + \frac{|H_1\left(\nu\right)|^2s_X\left(\nu\right)}{|H_2\left(\nu\right)|^2s_W\left(\nu \right)} \big).
\end{align}
Similarly, we can also show that 
\begin{align} \label{MI_3D_RII}
 \I\big(\big\{d\widetilde{X}_R\left( \nu\right),d\widetilde{X}_I\left( \nu\right) \big\}; & d\widetilde{Y}_I\left( \nu\right)\big) = \nonumber \\
 &  \frac{1}{2}\log \big(1 + \frac{|H_1\left(\nu\right)|^2s_X\left(\nu\right)}{|H_2\left(\nu\right)|^2s_W\left(\nu \right)} \big).
\end{align}
From \eqref{Spec_comp_independent}, \eqref{MI_3D_RIR} and \eqref{MI_3D_RII}, we have
\begin{align}
\MI_{XY}\left(\nu,\nu\right) & = 2\times\I\big(\big\{d\widetilde{X}_R\left( \nu\right),d\widetilde{X}_I\left( \nu\right) \big\}; d\widetilde{Y}_R\left( \nu\right) \big) \nonumber \\
& =  \log \big(1 + \frac{|H_1\left(\nu\right)|^2s_X\left(\nu\right)}{|H_2\left(\nu\right)|^2s_W\left(\nu \right)} \big).
\end{align}
At $\nu = 0$, MI-in-frequency between $X$ and $Y$ is equal to $ \I\big(\big\{d\widetilde{X}_R\left( \nu\right),d\widetilde{X}_I\left( \nu\right) \big\}; d\widetilde{Y}_R\left( \nu\right)\big)$, since the imaginary part of $Y$ is zero.

\paragraph{Relationship between MI in frequency and coherence} The coherence $C_{XY}\left( \nu\right) \in \left[0,1 \right]$ between two processes $X$ and $Y$ related by \eqref{mvar_cont_eq} is given by
\begin{align}
C_{XY}\left( \nu\right) = \frac{|s_{XY}\left(\nu\right)|^2}{s_X\left(\nu\right) s_Y\left( \nu\right)} & = \frac{|H_1\left(\nu\right)|^2s_X\left(\nu\right)}{s_X\left(\nu\right)|H_1\left(\nu\right)|^2+s_W\left(\nu\right)|H_2\left(\nu\right)|^2}. \nonumber
\end{align}
\begin{align}
\Rightarrow -\log\left(1-C_{XY}\left(\nu\right)\right) & = \log \big(1 + \frac{|H_1\left(\nu\right)|^2s_X\left(\nu\right)}{|H_2\left(\nu\right)|^2s_W\left(\nu \right)} \big) \nonumber \\
& = \MI_{XY}\left(\nu,\nu\right). \label{MI_in_freq_coherence}
\end{align}

\paragraph{Proof of Theorem~\ref{theorem4}} Now we consider two discrete-time Gaussian stochastic processes $X\left[n\right]$ and $Y\left[n\right]$ that are related by 
\begin{align} \label{mvar_discrete}
y[n] = h_1[n] \ast x[n] + h_2[n] \ast w[n],
\end{align}
where $h_1[n]$ and $h_2[n]$ are the impulse responses of two discrete-time linear, time-invariant filters. \eqref{mvar_discrete} is the discrete-time equivalent of \eqref{mvar_cont_eq}. It was shown in chapter 10 in \cite{pinsker1960} that mutual information between the discrete-time Gaussian stochastic processes $X\left[n\right]$ and $Y\left[n\right]$ is related to coherence according to 
\begin{align} \label{MI_coherence}
\I\left(X;Y\right) = -\int\limits_{0}^{0.5} \log \left(1-C_{XY}\left(\lambda\right)\right) d\lambda.
\end{align}
From \eqref{MI_in_freq_coherence} and \eqref{MI_coherence}, we have 
\begin{align}
\I \left(X;Y\right) = \int\limits_{0}^{0.5} \MI_{XY}\left(\lambda,\lambda\right) d\lambda.
\end{align}

\end{document}